\documentclass[twocolumn,aps]{revtex4}
\usepackage{amsmath,graphicx}

\begin{document}
\title{Electromagnetic Force and the Maxwell Stress
Tensor in Condensed Systems}
\author{Mario Liu~\cite{email}
\\ Institut f\"{u}r Theoretische Physik, Universit\"{a}t Hannover,
\\30167 Hannover, Germany, EC\\ and\\ Klaus Stierstadt
\\ Sektion Physik, Universit\"at M\"{u}nchen,
\\ 80539 M\"{u}nchen, Germany, EC}
\date{\today}

\begin{abstract}
While the electromagnetic force is microscopically simply
the Lo\-rentz force, its macroscopic form is more
complicated, and given by expressions such as the Maxwell
stress tensor and the Kelvin force. Their derivation is
fairly opaque, at times even confusing, and their range
of validity all but a well kept secret. These
circumstances unnecessarily reduce the usefulness and
trustworthiness of some key quantities in macroscopic
electrodynamics.

This article presents a thorough yet pedagogical
derivation of the Max\-well stress tensor and
electromagnetic force in condensed media. It starts from
universally accepted inputs: conservation laws,
thermodynamics and the Maxwell equations. Simplifications
are considered for various limits, especially the
equilibrium, with a range of validity assigned to each
expression. Some widespread misconceptions are
scrutinized, and hidden ambiguities in popular notations
revealed.

A number of phenomena typical of strongly polarizable
systems, especially ferrofluid, are then considered. In
addition to enhancing the appreciation of these systems,
it helps to solidify the grasp of the introduced concepts
and derived formulas, and it demonstrates the ease with
which the Maxwell stress tensor can be handled, inviting
theorists and experimentalists alike to embrace this
useful quantity.
\end{abstract}\maketitle

\tableofcontents
\section{Introduction}
\subsection{The Coarse-Grained Force}
Studying polarizable and magnetizable materials, one
central question concerns the electromagnetic force on
matter. This is especially true for liquids (such as
ferrofluids~\cite{rz,shli,blums}), as these respond not
only to the total force, but also to the spatially
varying force density. In principle, one may take the
Lorentz force as an exact microscopic expression and
coarse-grain it to obtain the macroscopic
force~\cite{fd},

\begin{equation}\label{a1}
\langle{\bf f}_{\rm L}\rangle=\langle\rho_e{\bf
E}\rangle+\langle\mbox{\boldmath$j$}_e\times{\bf
B}\rangle.
\end{equation}
Unfortunately, this is of little practical value, as we
neither have, nor indeed are interested in, the detailed
microscopic information of the fields $\rho_e,
\mbox{\boldmath$j$}_e, {\bf E}, {\bf B}$. On the other
hand, although we do have the knowledge of the
coarse-grained fields $\langle\rho_e\rangle$,
$\langle\mbox{\boldmath$j$}_e \rangle$, $\langle{\bf
E}\rangle$ and  $\langle{\bf B}\rangle$, or may obtain
them via the macroscopic Maxwell equations, it is clear
that the difference between the above true expression and
the ``fake" Lorentz force,

\begin{equation}\label{a2}
\langle\rho_e\rangle \langle{\bf E}\rangle+ \langle
\mbox{\boldmath$j$}_e\rangle \times \langle{\bf
B}\rangle,
\end{equation}
may become quite important: Think of an iron nail in a
magnetic field, which is obviously subject to an
electromagnetic force, although the ``fake" force
vanishes, as $\langle\rho_e\rangle, \langle
\mbox{\boldmath$j$}_e\rangle\to0$.

The difference between the true and fake Lorentz force is
frequently taken to be the Kelvin force, being
$P_i{\boldsymbol\nabla}\langle E_i\rangle$ in the
electric case~\cite{LL8}, where $P_i$ denotes the
polarization, with summation over repeated indices
implied. Originally, the formula was microscopically
derived: First write down the force exerted by an
external field on a dipole, eg by taking the derivative
of the electrostatic energy with respect to the position
of the dipole for given field. Then, assuming that the
dipoles are too far apart to interact, and to feed back
to the given field, take the total force density as
simply the sum of all forces exerted on the dipoles in a
unit volume. Clearly, this result neglects nonlinear
terms in $P_i$ (which account for interaction) and is
valid only in the dilute limit. This implies small
polarization and therefore weak force, and one is free to
add a second order term, writing the Kelvin force as
\begin{equation}\label{a3}
P_i{\boldsymbol\nabla} (\langle E_i\rangle
+P_i/\epsilon_0)\equiv P_i {\boldsymbol\nabla}
D_i/\epsilon_0.
\end{equation}
There is a more sophisticated macroscopic derivation of
the Kelvin force that we shall discuss next. Stating that
the same constraint applies here, and one may still write
the Kelvin force either as $P_i{\boldsymbol\nabla}\langle
E_i\rangle$ or as $P_i {\boldsymbol\nabla}
D_i/\epsilon_0$ is bound to raise a few eyebrows. But as
will be shown in this review, it is really a rather
straight-forward implication of the assumptions made
during its derivation. By the same token, for linear
constitutive relation, the Kelvin force is only valid for
small susceptibility, $\chi\ll1$, to linear order in
$\chi$.

The same holds for the magnetic case, where the
corresponding force is much larger in praxis (as it is
not subject to ionization or residual conductivity). The
Kelvin force expression is again valid only when weak, to
linear order in the magnetization $M\ll\langle B\rangle$,
or the magnetic susceptibility $\chi^{\rm m}\ll1$. And
one may take its expression either in the usual
form~\cite{rz,LL8}, as

\begin{equation}\label{a4}
M_i{\boldsymbol\nabla} (\langle
B_i\rangle-\mu_0M_i)\equiv\mu_0M_i{\boldsymbol\nabla}
H_i,
\end{equation}
or equally justified, as $M_i{\boldsymbol\nabla}\langle
B_i\rangle$. Consequently and remarkably, the Kelvin
force is not usually a valid formula in ferrofluids,
where frequently $M\approx H$ and $\chi^{\rm m}\approx1$
--- though it is widely employed there.

The nowadays generally accepted method to derive the
electromagnetic, or ``ponderomotive" force  $f^{\rm
P}_i$, cf \S15 of {\em Electrodynamics in Continuous
Media} by Landau and Lifshitz~\cite{LL8}, is a definit
improvement over the microscopic one given above. It is
completely macroscopic and starts from identifying
$f^{\rm P}_i$ as

\begin{equation}\label{a5}
f^{\rm P}_i=-\nabla_j[\Pi^{\rm tot}_{ij}
-P(T,\rho)\delta_{ij}],
\end{equation}
where $-\Pi^{\rm tot}_{ij}$ denotes the macroscopic
Maxwell stress tensor, and $P(T,\rho)$ the zero-field
pressure -- the pressure that is there for given
temperature and density, but without the external
field. The ponderomotive force $f^{\rm P}_i$ reduces
to the Kelvin force, if the polarization and
magnetization are (assuming the dilute limit)
proportional to the density $\rho$.

The idea behind the identification seems attractively
simple: Since $-\nabla_j\Pi^{\rm tot}_{ij}$ is the total
force on a volume ele\-ment, and $-\nabla_iP$ the force
at zero field, their difference $f^{\rm P}_i$ must be the
force due to the external field. Unfortunately, as we
shall show in this review, the resultant force is rather
ambiguous. The reason lies with the zero-field pressure
$P(T,\rho)$, which is not a unique quantity, because its
value depends on the chosen variables, eg
$P(T,\rho)\not=$ $P(s,\rho)$, where $s$ is the entropy
density. The conventionally defined Kelvin force is valid
only under isothermal conditions with the density kept
constant. Under adiabatic conditions, with $s$ given, the
temperature $T$ is known to change with the field, so
will $P(T,\rho)$, the alleged zero-field pressure. Any
field dependence implies that $P(T,\rho)$ contributes to
the electromagnetic force. Employing the force of
Eq~(\ref{a5}) alone then leads to wrong predictions.

In physics, we tend to start from a robust base of broad
applicability when studying a given situation, narrowing
our focus along the way only when necessary. Clearly,
while the universally valid Lorentz force is one such
base, the fragile ponderomotive force, and even more
specific, the Kelvin force are not. This makes it
desirable to look for better alternatives, one of which
is employing the Maxwell stress tensor directly.

\subsection{The Maxwell Stress Tensor}\label{MST}
Given in terms of thermodynamic variables, the Max\-well
tensor is an unequi\-vo\-cally macroscopic,
coarse-grained quantity. It holds for all conceivable
systems and is subject only to the validity of local
equilibrium -- a constraint in frequency but not in the
type and strength of interaction. Besides, more
frequently than not, in understanding the many facets of
strongly polarizable systems, it is more convenient to
employ the Maxwell tensor directly, rather than to follow
the detour via pressure and the ponderomotive force
$f^{\rm P}_i$ --- in spite of having to deal with
quantities such as the chemical potential or the entropy.

Given the Maxwell tensor's pivotal role for a solid
understanding of electromagnetism in condensed systems,
(not least as the starting point for forces), it is
especially unfortunate that one is hard pressed to find a
fully convincing derivation of its form. The classic
reference is again to the otherwise excellent book by
Landau and Lifshitz~\cite{LL8}, the relevant \S15 of
which, however, leaves many readers unconvinced, even
confused. The reasons for this are presented in the four
sections below.

\subsubsection{Three Obvious Objections} The three more
obvious reasons for the frustration of the readers are
(i) the gradient terms, (ii) the magnetic field
dependence, and (iii) the concentration: First, the
Maxwell tensor is derived considering a capacitor. Its
geometry is so simple that terms proportional to the
gradients of temperature and field may have been missed.
Although these are asserted to be negligible in a
footnote, one feels the need for convincing arguments.

Second, the magnetic terms are obtained, in \S 35 of
~\cite{LL8}, by the replacement ${\bf E\to H}$, ${\bf
D\to B}$. This would be justified if the system under
consideration does not contain sources, and the
respective Maxwell equations are the same,
${\boldsymbol\nabla}\!\times\!{\bf E}=0$,
${\boldsymbol\nabla}\!\cdot\!{\bf D}=0$ versus
${\boldsymbol\nabla}\!\times\!{\bf H}=0$,
${\boldsymbol\nabla}\!\cdot\!{\bf B}=0$. In the geometry
used to derive the electric part of the stress tensor,
however, the charge on the capacitor plates is essential
for producing the electric field. Therefore, an
independent consideration for the magnetic part is
necessary, with a current carrying coil replacing the
charge laden capacitor. In \S 4 of his widely read book
on ferrofluids~\cite{rz}, Rosensweig points to this gap,
and aims to fill it. Unfortunately, the calculation
starts from a faulty assumption, discrediting the
derivation, cf section~\ref{mag}.

Finally, neither Landau and Lifshitz nor Rosen\-sweig
study the concentration dependence of the stress
tensor, although it is an important thermodynamic and
hydrodynamic variable in ferrofluids, liable to
undergo much greater variations than the density, and
hence of experimental importance.

\subsubsection{The Range of Validity} Next, the lack of
any discussion on the range of validity for the obtained
expression further undermines the trust of the readers.
The derivation is thermodynamic in nature and takes place
in a stationary capacitor, so the result is apparently
valid only in equilibrium and for vanishing liquid
velocity. Consistent to this, the static Maxwell
equations for the rest frame are repeatedly employed when
deriving the electromagnetic force from the Maxwell
tensor.

Nevertheless, it is customary to take for granted that
the form of the Maxwell tensor and the electromagnetic
force thus derived also hold in dynamic situations, at
finite frequencies and nonvanishing liquid velocity. This
may well be approximately true for some circumstances,
but which and why? Should we not rather derive
dynamically valid expressions for the Maxwell stress
tensor and the electromagnetic force, and justify the
employed approximations by estimating the magnitude of
the terms neglected?

\subsubsection{A Technical Flaw} Then there is a technical
flaw that actually renders the derivation mathematically
invalid: There is no question that one can always obtain
a tensorial quantity $\Pi^{\rm tot}_{ij}$ by evaluating
the scalar $\Pi^{\rm tot}_{ij}\xi_in_j$ while varying the
unit vectors $\xi_i$ and $n_j$. Taking each of the two
unit vectors to successively point in all three
directions, we obtain nine different results for
$\Pi^{\rm tot}_{ij} \xi_in_j$, each a component of
$\Pi^{\rm tot}_{ij}$. In~\cite{LL8}, $n_j$ is the surface
normal, $\xi_i$ the direction of the virtual
displacement, and the Maxwell tensor $-\Pi^{\rm
tot}_{ij}$ is denoted as $\sigma_{ij}$.

With the electric field being the only preferred
direction, $\xi_i$ and $n_j$ need to be both along and
perpendicular to the field. In \S 15 of~\cite{LL8}, only
the displacement of the capacitor plate is considered,
implying that the surface normal $n_j$ is kept parallel
to the E-field throughout. Therefore, if the direction of
the electric field is taken as $\bf\hat e_x$, only the
components $\Pi_{xx}$, $\Pi_{yx}$ and $\Pi_{zx}$ could
have been evaluated. A footnote states: {\em It is not
important that in this derivation $\bf E$ is parallel to
{\bf n}, since $\Pi_{ij}$ can obviously depend only on
the direction of $\bf E$, not on that of $\bf n$.} This
is incorrect, because although $\Pi_{ij}$ does not depend
on {\bf n}, the scalar $\Pi_{ij}\xi_in_j$ does: Take for
instance the term $D_iE_j$, which is part of the Maxwell
stress, and which indeed does not depend on $\bf n$. Yet
with $\bf D,E\ \|\ e_x$, we have $D_iE_jn_j\not=0$ for
$\bf n\ \|\ e_x$, and  $D_iE_jn_j=0$ if $\bf n\perp e_x$.

\subsubsection{Conceptual Problems} Finally, there are
conceptual problems of more fundamental nature. The
Maxwell stress tensor is introduced in \cite{LL8} by the
terse sentences: {\em It is well known that the forces
acting on any finite volume in a body can be reduced to
forces applied to the surface of that volume. This is a
consequence of the law of conservation of momentum.} The
surface force density is subsequently taken as the stress
tensor.

Momentum conservation is without further elaboration a
confusing argument, because the very idea of force is
something that alters the momentum of particles. Being
accelerated by external fields, the momentum of a lump of
material is not conserved, and there is no reason why the
momentum density $\rho\mbox{\boldmath$v$}$ should satisfy
a continuity equation. On the other hand, what the
electromagnetic field does is to {\em exchange} momentum
with the material, so the total momentum of both remains
conserved, which is, as we shall see,
$\mbox{\boldmath$g$}^{\rm tot} =
\rho\mbox{\boldmath$v$}+{\bf E}\times{\bf H}$. This
momentum density does satisfy a continuity equation,
$\dot g_i^{\rm tot}+ \nabla_k\Pi_{ik}^{\rm tot} = 0$.
(Even in the presence of fields that accelerate
particles, the total system -- including particles and
the field producing charge carriers -- is still
translational invariant, with the associated total
momentum a locally conserved quantity, see \S 75
of~\cite{LL8}, Ch 14 of~\cite{dGM}, and~\cite{henjes}.)
The negative of the associated momentum flux,
$-\Pi_{ik}^{\rm tot}$, is the properly defined,
dynamically valid Maxwell tensor. It holds its ground for
finite frequencies, nonvanishing velocities of the
medium, and including dissipations. Understanding the
conservation of $\boldsymbol g^{\rm tot}$
$=\rho\mbox{\boldmath$v$}$ $+{\bf E}\times{\bf H}$
renders the derivation of the Maxwell tensor cogent and
coherent, even though the second term is, under usual
circumstances in condensed matter, much smaller than the
first.

\subsection{Content of this Article}
Our aim in this paper is a careful and rigorous
derivation of the Maxwell tensor and the dynamically
valid electromagnetic force, in a presentation that is
detailed and pedagogical. We also consider applications,
especially for experimentally relevant situations in
ferrofluids, to give hands-on illustrations on how to
employ these quantities.

The paper may be divided into two parts, The first
consists of Chapter \ref{gene} and concentrates on the
fundamental concepts. The hydrodynamic equations for a
dense system exposed to electromagnetic fields ---
including especially the form of the Maxwell tensor ---
are derived here. The second part (Chapter \ref{gene2}
and \ref{exp}) is more practically concerned. It starts
from the Maxwell tensor, derives the electromagnetic
force, considers different simplifications, and studies a
number of experimentally relevant situations. Although we
believe that Chapter \ref{gene} provides useful and
necessary insights, the second part could if necessary be
read alone.

Chapter \ref{gene} starts by differentiating two
types of theories: First, the high-resolution theory
for a low density system, with at most one particle
per infinitesimal volume element, or per grain (as in
photographs); second, the low resolution theory for a
high density system, with many particles per grain.
The microscopic Maxwell equations and the Newtonian
equation of motion (including the Lorentz force)
belong to the first type. There are no hidden
charges, polarization or magnetization here, and we
know the whereabouts of every single particle. The
second type is represented by the macroscopic Maxwell
equations, or any thermodynamic and hydrodynamic
theories. The question about the coarse-grained form
of the electromagnetic force arises here.

We shall consider three systems with increasing
densities, to be accounted for by theories of
decreasing resolution and increasing complexities:
The first system (section~\ref{high}) is a dilute gas
of charged particles, which is well accounted for by
the Newtonian equation of motion and the microscopic
Maxwell equations. The next system
(section~\ref{part}) is a slightly dissociated liquid
of particles possessing negligible electric and
magnetic dipole moments. So there are many particles,
but at most one charge carrier, per grain. The
appropriate theory here is of a mixed type, being a
combination of the hydrodynamic theory and the
microscopic Maxwell theory. The former accounts for
all particles, the latter accounts for the spatially
slowly varying field, either externally imposed or
produced by the few charge carriers. Finally, in
section~\ref{EoM}, we consider a dense system,
including dipole moments and hidden sources. It is to
be accounted for by the genuinely low-resolution,
hydrodynamic-type Maxwell theory that is our goal to
derive and consider in details.

In all these systems, there are a conserved energy
$u^{\rm tot}$ and a conserved momentum density,
$g_i^{\rm tot}\not=\rho v_i$, both consist of
material and field contributions. These we shall
choose as two of our dynamic variables, with the
equations of motion $\dot u^{\rm tot}+
\nabla_kQ_k^{\rm tot} = 0$ and $\dot g_i^{\rm tot}+
\nabla_k\Pi_{ik}^{\rm tot} = 0$, where
$-\Pi_{ik}^{\rm tot}$ is the Maxwell stress tensor.
We shall obtain explicit expressions for $g_i^{\rm
tot}$ and $\Pi_{ik}^{\rm tot}$ in all three systems
--- whereby finding the expressions in
the genuinely dense, third medium is rendered very
much simpler, and more transparent, by the
considerations in the previous two dilute systems.

In Chapter~\ref{gene2}, starting from the Maxwell tensor,
we shall first obtain the general expression for the
electromagnetic force density, valid for arbitrarily
strong fields, moving medium and finite frequencies,
given essentially by $-(s{\boldsymbol \nabla} T+\rho
{\boldsymbol\nabla}\xi)$, where $\xi$ is the chemical
potential. Although this expression is already valid
without field, it is the proper formula for the quantity
usually written as the sum of the zero-field pressure
gradient and the ponderomotive force. In contrast to the
latter, however,  $-(s{\boldsymbol \nabla} T+\rho
{\boldsymbol\nabla}\xi)$ is valid for any sets of
variables, and for arbitrarily large polarization and
magnetization.

In equilibrium, since both the temperature $T$ and
the chemical potential $\xi$ are uniform, this force
density is identically zero. Nevertheless, electric
and magnetic fields remain operative, may lift a
polarizable system off the ground, deform its shape
and generally alter its hydrostatics --- all signs
that forces of electromagnetic origin are at work. In
section \ref{surfF}, the responsible equilibrium
force is identified as a surface force. Its form
follows directly from the simplifications that occur
for the Maxwell stress if the system is in
equilibrium (section \ref{MSE}). As most of the
experiments considered in Chapter \ref{exp} reflect
varying appearances of the equilibrium force, this is
a useful and concise result.

A thermodynamic rederivation of the Maxwell stress
tensor is given in section~\ref{ener}, an approach
which is similar in spirit to that of Landau and
Lifshitz~\cite{rz,LL8}, yet carefully freed of the
flaws listed in section \ref{MST} above. In
comparison to the derivation given in Chapter
\ref{gene}, this here is narrower in focus and
validity, but simpler. Chapter~\ref{gene2} ends with
section~\ref{MN}, which points out the inherent
ambiguity of the zero-field pressure $P(T,\rho)$, as
it is independent of external fields only if $T$ and
$\rho$ are kept constant. We stress this point by
repeating the separation of the zero-field pressure
from the Maxwell stress for various combinations of
variables, obtaining many forms of the ponderomotive
force, all rather different from each other.
Furthermore, a warning is issued on the unsuspected,
narrow range of validity for various expressions in
the notation containing the zero-field pressure,
especially the Maxwell stress in the approximation
$(P+\frac{1}{2}H^2)\delta_{ij}-H_iB_j$, and the
Kelvin force. Neither is valid for large
magnetization and magnetic susceptibility, $M\approx
H$ and $\chi^{\rm m}\approx1$.

The very useful magnetic Bernoulli equation by
Ro\-sen\-sweig~\cite{rz} is unfortunately also given in
this notation. An formulation free of ambiguity is found,
along with the generalization to include the variation of
concentration.

In Chapter~\ref{exp}, we study a number of phenomena
typical of strongly polarizable systems. Most of
these experiments have already been given detailed
theoretical analyses, especially by Rosensweig in his
comprehensive book on ferrofluids~\cite{rz}. The
reason to repeat them here is threefold: First, to
demonstrate that it is not necessary to introduce the
zero field pressure, that experimental phenomena may
be well and easily accounted for employing the
Maxwell stress directly, using a generally valid,
unambiguous notation. Second, we believe that our
derivations are well streamlined and easy to follow.
This is achieved by stressing the simplification of
equilibrium, and by consistently using only the
Maxwell stress and the associated boundary
conditions. In contrast, confusingly many pressures
are deemed necessary in~\cite{rz}. Finally, we aim to
amend some of the physics and especially include the
concentration of magnetic particles as an independent
variable. Being inhomogeneous where the field is, the
concentration alters experimental outcomes and is
something theoretical considerations need to include.

Since mesoscopic particles diffuse slowly, the
equilibrium between concentration and field gradient
needs time (up to days, even weeks) to establish. So
for times much briefer, the concentration will remain
constant and may indeed be taken as dependent. On the
other hand, strong gradients, small geometry, large
particles will all considerably shorten the
equilibrating time (down to fractions of a minute),
then the spatially varying concentration will be an
important feature in any system. Besides, long
contacts with field gradients are also of interest
(eg in magnetic O-rings). Although the non-uniform
equilibrium distribution of magnetic particles is not
a well studied subject, their diffusion under
enforced non-equilibrium situation is,
see~\cite{blums,odenbach}.

Now a rundown of the considered experiments. In
section~\ref{denva}, we study the variation of
density and concentration in the presence of
inhomogeneous fields: While the variation of density
is a measured phenomenon, that of the concentration
is not -- although its variation is much more
pronounced, by many orders of magnitude. A vertical,
current-carrying wire that goes through a dish
containing ferrofluid will drag up the ferrofluid
along the wire. Our calculation in section~\ref{vccw}
includes the surface tension and the variation of
concentration. The following section~\ref{levi}
considers how far an external field will raise the
level of a fluid column, and what a pressure gauge
actually measures within the ferrofluid that is
subject to external fields. Section~\ref{scrap} deals
with rather similar physics and considers magnetic
O-rings and scrap separation. Section~\ref{elli}
contains a simple calculation on the elliptical
deformations a droplet of ferrofluid undergoes along
the external magnetic field.

\subsection{Two Caveats}
There are two points we shall not address in this
article. First, we shall not consider any dissipative
terms, although the framework presented here is well
suited for including them. This was a difficult
decision, as the dissipative terms account for
irreversibility, without which no macroscopic,
coarse-grained theory is complete, and because they
give rise to additional electromagnetic forces of
comparable magnitude and experimental relevance. Yet,
including these terms would have considerably
lengthened an already lengthy article, and further
complicated its formulas and arguments. We pledge
amendments for the future, and for now direct the
impatient readers to the
literature~\cite{shli,blums,hydM}.

Second, given the long and tortuous history of the
struggle to come to terms with the macroscopic
electromagnetic force, time and again forcing us to back
up from blind alleys, any attempt by us on a
comprehensive citation would bear historic rather than
scientific interests. As we see it, the consistent
thermodynamic treatment given first by Landau and
Lifshitz~\cite{LL8}, and later generalized to include
nonlinearity by Rosensweig and others~\cite{rz}, cf the
excellent summary by Byrne~\cite{byrne}, is so much more
superior than the many treatises preceding them, that
this represents the only worthwhile starting point for
any further considerations --- and consequently these two
are also the only ones subject to scrutiny and criticisms
here.

\subsection{Notes on Units} This paper will be in the SI
units throughout, though with a little twist to render
the display and manipulation of the formulas simple. We
define, and from now on shall extensively use, the fields
and sources
\begin{eqnarray}\label{hat}
{\bf H}={\bf \hat H}\sqrt{\mu_0},\quad {\bf B} ={\bf
\hat B}/\sqrt{\mu_0},\quad \varrho_e=
\hat\varrho_e/\sqrt{\epsilon_o},\nonumber\\ {\bf
E}={\bf \hat E}\sqrt{\epsilon_o},\qquad \,{\bf
D}={\bf \hat D}/\sqrt{\epsilon_o},\ \quad q=\hat
q/\sqrt{\epsilon_o}, \\ {\bf P\equiv D-E=\hat
P}/\sqrt{\epsilon_o},\quad {\bf M\equiv B-H=\hat
M}\sqrt{\mu_0}, \nonumber
\end{eqnarray}
where (again starting from here) the quantities with hats
are the usual ones, denominated in MKSA, or the SI-units.
The advantage is, all new fields have the same dimension,
$\sqrt{\rm J/m^3}$, and sensibly, $\bf H=B$ and $\bf D=E$
in vacuum. (Electric charge $q$ has the dimension
$\sqrt{\rm Jm}$, and its density $[\varrho_e]= \sqrt{\rm
J/m^5}$.) Written in the new fields, all formulas are rid
of the ubiquitous (and content-free) factors $\epsilon_0$
and $\mu_0$. To return to the more familiar MKSA-fields
{\em at any instance}, simply employ Eqs~(\ref{hat}).
(These fields are sometimes referred to as being in the
Heaviside-Lorentz, or the rational, units. But saying so
immediately leads to heated and pointless discussions. We
stick to the SI units throughout and use the above fields
only as a convenient shorthand that may be abandoned
whenever one chooses.)

Some literature, notably~\cite{LL8}, employs cgs, or {\em
Gauss}-units. To ease comparison with them, simply note
that the {\em Gauss}-fields, with tilde, may be obtained
any time via
\begin{eqnarray}\label{tilde}
{\bf H}={\bf \tilde H}/\sqrt{4\pi},\quad {\bf B} ={\bf
\tilde B}/\sqrt{4\pi},\quad \varrho_e=
\sqrt{4\pi}\,\tilde{\varrho_e},\nonumber\\ {\bf E}={\bf
\tilde E}/\sqrt{4\pi},\,\quad{\bf D}={\bf \tilde
D}/\sqrt{4\pi},\quad j_e=\sqrt{4\pi}\,\tilde{j_e},
\\ {\bf P}=\sqrt{4\pi}{\bf \tilde P},\qquad {\bf
M}=\sqrt{4\pi}\,{\bf \tilde M}, \nonumber
\end{eqnarray}

\section{Hydrodynamic Theories}\label{gene}
\subsection{Dilute Plasma}\label{high}
We first consider a
rarefied system of charged particles, and choose a
resolution that is high enough that each grain (or
infinitesimal volume element) contains one or less
particles. (This theory is meant as a starting point,
to clarify a few concepts important for the more
complex theories of the following chapters. So we
shall simply discard the possibility that even in a
rarefied gas, two particles will occasionally come
close to each other.) The microscopic Maxwell
equations account for the time evolution of the
electromagnetic field for given sources,
\begin{eqnarray} \label{1}
{\bf{\boldsymbol\nabla} \cdot e}=\rho_e,
\quad\qquad{\bf{\boldsymbol\nabla} \cdot b}= 0,  \\
\label{1a}{\bf\dot e}=c{\boldsymbol\nabla}\times{\bf
b}-\rho_e\mbox{\boldmath$v$},\quad{\bf \dot
b}=-c{\boldsymbol\nabla}\times{\bf  e},
\end{eqnarray}
while the feedback of the field on the motion of the
sources is given by the Newtonian equation of motion,
\begin{equation}\label{2}
m\dot{\mbox{\boldmath$v$}}^\alpha=q({\bf
e}+\mbox{\boldmath$v$}^\alpha\times{\bf b}/c),
\end{equation}
one for each particle $\alpha$. (To emphasize the
fact that we are here dealing with the
high-resolution, microscopically accurate fields,
these are denoted by the lower case letters ${\bf
e\equiv\hat e}\sqrt{\epsilon_o},\ {\bf b\equiv\hat
b}/\sqrt{\mu_0}$, while the coarse-grained fields
below are denoted by the usual $\bf D, E, H, B$.)
Eqs~(\ref{1},\ref{1a},\ref{2}) represent a
conceptually simple and complete theory, but it
contains a notational inconsistency: The Maxwell
equations are an Euler type theory, accounting for
the time evolution of fields at a given point in
space, while the Newtonian equation is of the
Lagrange type, which concentrates on a given
particle. (So the term $\rho_e{\boldsymbol v}$  in
Eq~(\ref{1a}) denotes the electric current at a space
point, while ${\boldsymbol v}^\alpha$ in Eq~(\ref{2})
is the velocity of a particle.) As only the Euler
formulation lends itself to a scaling-up of the
grains and a reduction of the resolution, we shall
first find the Euler formulation for the Newtonian
equation, before using it to draw a number of
conclusions useful for dense systems.

\subsubsection{Euler Version of Newtonian
equation\label{eule}} Since there is at most one particle
per grain, of volume ${\rm V_G}$, we may identify the
velocity, mass and charge of a volume element with that
of the particle occupying it at a given instance, and
take all three to be zero if there is no particle. As a
result, we obtain three (highly discontinuous) fields:
$\mbox{\boldmath$v$}^\alpha\to \mbox{\boldmath$v$}({\bf
r},t)$, $m/{V_G}\to\rho({\bf r},t)$,
$q/{V_G}\to\rho_e({\bf r},t)$. The many Newtonian
equations of motion then reduce to one field equation,
\begin{equation}\label{3}
\rho\textstyle{\frac{{\rm d}}{{\rm
d}t}}\mbox{\boldmath$v$}\equiv
\rho[\dot{\mbox{\boldmath$v$}}+ (\mbox{\boldmath$v$}\cdot
{\boldsymbol\nabla})\mbox{\boldmath$v$}] =\rho_e({\bf
e}+\mbox{\boldmath$v$}\times{\bf b}/c).
\end{equation}
(The quantity $\nabla v$ is to be taken from the
velocity of the same particle at two successive
moments.) It is now essential to explicitly include
the continuity equation,
\begin{equation}\label{4}
\dot\rho+{\boldsymbol\nabla}\cdot
(\rho\mbox{\boldmath$v$})=0,\end{equation}
that in the Lagrange version is implicit, nearly
incidentally contained in the fact that one does not
loose any of the many equations (\ref{2}). [The
continuity equation for the charge density $\rho_e$ is
implied by the Maxwell equations Eq~(\ref{1},\ref{1a}).]

The material contributions to the energy and momentum
density are, respectively,
\begin{equation}\label{5}
u^{\rm M}=\rho(c^2+v^2/2),\quad
\mbox{\boldmath$g$}^{\rm M}=\rho\mbox{\boldmath$v$}.
\end{equation}
where the first comprises of the rest energy and the
(non-relativistic) kinetic energy. Employing
Eqs~(\ref{3},\ref{4}), we find
\begin{eqnarray}\label{7}
\dot u^{\rm M}+{\boldsymbol\nabla}\cdot{\bf Q}^{\rm
M}=\rho_e(\mbox{\boldmath$v$}\cdot{\bf e}),\\ \dot
g_i^{\rm M}+\nabla_k\,\Pi^{\rm M}_{ik}= \rho_e({\bf
e}+\mbox{\boldmath$v$}\times{\bf b}/c)_i,
  \label{8}\\
\label{9}{\bf Q}^{\rm M}= u^{\rm
M}\mbox{\boldmath$v$},
\quad \Pi^{\rm M}_{ik}=g^{\rm M}_iv_k.
\end{eqnarray}

\subsubsection{The Field Contributions\label{fiel}}
The field contribution to the energy and momentum
density are
\begin{equation}\label{11}
u^{\rm F}=\textstyle{\frac{1}{2}}(e^2+b^2),\quad
\mbox{\boldmath$g$}^{\rm F}= {\bf e\times b}/c.
\end{equation}
From the Maxwell equations
(\ref{1},\ref{1a}) we deduce
\begin{eqnarray}\label{12}
\dot u^{\rm F}+{\boldsymbol\nabla}\cdot{\bf Q}^{\rm
F}=-\rho_e(\mbox{\boldmath$v$}\cdot{\bf e}),\quad {\bf
Q}^{\rm F}=c\,{\bf e\times b}, \\ \dot g_i^{\rm
F}+\nabla_k\Pi_{ik}^{\rm F}=-\rho_e({\bf
e}+\mbox{\boldmath$v$}\times{\bf b}/c)_i, \label{13}\\
\Pi^{\rm F}_{ik}=( e^2+b^2 -u^{\rm F}) \delta_{ik} -
e_ie_k-b_ib_k. \label{15}
\end{eqnarray}
Note the relationship $\mbox{\boldmath$g$}^{\rm F}= {\bf
Q}^{\rm F}/c^2$. This is far from accidental and derives
from the symmetry of the relativistic energy-momentum
4-tensor, $\Pi^{\rm F}_{\alpha\beta}= \Pi^{\rm
F}_{\beta\alpha}$, because $cg_k^{\rm F}=\Pi^{\rm
F}_{k4}$, $Q^{\rm F}_k/c=\Pi^{\rm F}_{4k}$. Less
formally, $\mbox{\boldmath$g$}^{\rm F}= {\bf Q}^{\rm
F}/c^2$ may also be seen as the conservation of the field
angular momentum for $\rho_e\to0$. The angular momentum
density $\mbox{\boldmath$\ell$}^{\rm F} \equiv{\bf r}
\times\mbox{\boldmath$g$}^{\rm F}$ is a locally conserved
quantity in neutral systems. Rewriting Eq~(\ref{13})
(with $\rho_e\to0$) as $\partial\ell_i^{\rm F}/\partial
t+\nabla_m(\epsilon_{ijk}r_j \Pi_{km}^{\rm F}) =
\epsilon_{ijk}\Pi_{kj}^{\rm F}$, we observe that the
angular momentum $\ell_i^{\rm F}$ satisfies a continuity
equation only if the stress tensor is symmetric. Although
this argument seems to require merely the symmetry of the
momentum 3-tensor, $\Pi^{\rm F}_{ik}= \Pi^{\rm F}_{ki}$,
we know that a nonvanishing $\Pi^{\rm F}_{k4}- \Pi^{\rm
F}_{4k}$ in one inertial system will foul up the symmetry
of the 3-tensor in other systems, as the antisymmetric
parts of any 4-tensors mix in a Lorentz transformation.
Yet angular momentum is conserved in every inertial
system.

Since this reasoning is so general, it also holds for the
material part, $\mbox{\boldmath$g$}^{\rm M}= {\bf Q}^{\rm
M}/c^2$. Hence the expression for the momentum density is
actually $\mbox{\boldmath$g$}^{\rm
M}=\rho\mbox{\boldmath$v$} [1+v^2/(2c^2)]$, cf
Eqs~(\ref{5},\ref{9}), though we are quite justified to
neglect the second term in the nonrelativistic limit.
Later, when we have no prior knowledge of the form of the
momentum density, we shall deduce it from that of the
energy flux, as angular momentum is also conserved in
dense systems.

We register the fact that while the expression for the
energy density $u^{\rm F}=\frac{1}{2} (e^2+b^2)$ is a
genuine input -- independent of, and in addition to, the
Maxwell equations, the formula $\mbox{\boldmath$g$}^{\rm
F}= {\bf e\times b}/c$ is not, since ${\bf Q}^{\rm F}$ is
given once $u^{\rm F}$ is.

\subsubsection{Energy and Momentum
Conservation}\label{cons} The preceding two sections
allow the simple and noteworthy conclusion that our
starting equations imply local conservation of total
energy, momentum and angular momentum even in the
presence of charges, $\rho_e \not=0$. Taking $u^{\rm tot}
\equiv u^{\rm F}+u^{\rm M}$ and $\mbox{\boldmath$g$}^{\rm
tot} \equiv\mbox{\boldmath$g$}^{\rm
F}+\mbox{\boldmath$g$}^{\rm M}$, we find
\begin{eqnarray}\label{16}
\dot u^{\rm tot}+{\boldsymbol\nabla}\cdot{\bf Q}^{\rm
tot}=0,\quad \dot g_i^{\rm tot}+\nabla_k\Pi_{ik}^{\rm
tot}=0,
\\\label{20} \Pi_{ik}^{\rm tot}=\Pi_{ki}^{\rm
tot}=\Pi_{ik}^{\rm F}+\Pi_{ik}^{\rm M},\quad {\bf
Q}^{\rm tot}={\bf Q}^{\rm F}+{\bf Q}^{\rm M},\\ {\bf
Q}^{\rm tot}/c^2=\mbox{\boldmath$g$}^{\rm
tot}\approx\rho \mbox{\boldmath$v$}+ {\bf e\times
b}/c. \label{18a}
\end{eqnarray}
These results have been collected here because local
conservation of these quantities is always true,
independent of the above derivation tailored to a
dilute and finely resolved system. So we may start
from them as input next.

\subsection{Weakly Dissociated Gas}\label{part} Now we
consider a dense macroscopic system in its
hydrodynamic regime: To the above gas of dilute
charge carriers we add a dense system of neutral
particles with vanishing electric and magnetic
dipole moments. One example of this composite system
is a slightly ionized gas with negligible electric
and magnetic susceptibility. This is still a
comparatively simple system, as the highly
resolving, vacuum Maxwell equations
(\ref{1},\ref{1a}) remain valid -- and with them
every single of the equations of section~\ref{fiel},
about the field contributions to energy and
momentum.  The equations in section~\ref{eule},
about the respective material contributions, must be
modified, as these will now be accounted for by
three hydrodynamic variables: the entropy density
$s$, and the coarse-grained mass density
$\langle\rho\rangle$ and velocity $ \langle
v\rangle$ -- all smooth and slowly varying fields.
(The coarse-graining brackets are dropped below to
retain the simplicity of display.)

\subsubsection{The Material Contributions}\label{mate} The
hydrodynamic theory of the material part of our system is
given by two continuity and two balance equations,
\begin{eqnarray}
\dot\rho+{\boldsymbol\nabla}\cdot(\rho\mbox{\boldmath$v$})=0,
\quad \dot s+{\boldsymbol\nabla}
\cdot(s\mbox{\boldmath$v$})=0, \label{26}\\ \dot u^{\rm
M}+{\boldsymbol\nabla}\cdot{\bf Q}^{\rm M}=
\rho_e(\mbox{\boldmath$v$}\cdot{\bf e}), \label{27}\\
\dot g_i^{\rm M}+\nabla_k\,\Pi_{ik}^{\rm M}= \rho_e({\bf
e}+ \mbox{\boldmath$v$} \times{\bf b}/c)_i, \label{28}\\
{\bf Q}^{\rm M}=(u^{\rm M}+P)\mbox{\boldmath$v$},\quad
\Pi_{ik}^{\rm M}=g^{\rm M}_iv_k+P\delta_{ik}.\label{add1}
\end{eqnarray}
For $\rho_e\to0$, these are the well known hydrodynamic
equations of a neutral, isotropic liquid~\cite{LL6}. For
finite $\rho_e$, the right sides of Eqs~(\ref{27},
\ref{28}) have the given form because the field source
terms of Eqs~(\ref{12}, \ref{13}) remain unchanged, and
because summing up the respective right sides must yield
nil, such that total energy and momentum are conserved,
as given in Eqs(\ref{16}). (Since $v$ actually denotes
$\langle v\rangle$ now, the above formulas presume the
validity of ${\bf\dot e}=c{\boldsymbol\nabla}\times{\bf
b}-\rho_e\langle\mbox{\boldmath$v$}\rangle$, rather than
the original Eqs~(\ref{1a}). This is justified, however,
because charge carriers do move with $\langle v\rangle$
in the absence of dissipation.)

Comparing ${\bf Q}^{\rm M}$ and $\Pi_{ij}^{\rm M}$
here to those of Eqs~(\ref{9}), the only apparent
difference is the appearance of the pressure $P$,
clearly an expression of the present system being
dense and interacting. However, there is more than
meets the eyes, and the most important one is our
ignorance about the explicit expression for the
energy $u^{\rm M}$. (Also, being a hydrodynamic,
coarse-grained theory, the equations of motion now
contain dissipative terms to account for
irreversibility and entropy production. As mentioned
in the introduction, however, we shall neglect these
in this article.)

In the co-moving, local rest frame of the liquid,
(denoted by the subscript $_0$,) the material energy
density in local equilibrium is given by the
thermodynamic expression,
\begin{equation}\label{21}
{\rm d}u^{\rm M}_0=T{\rm d}s+\xi_0{\rm d}\rho,
\end{equation}
which states the simple fact that $u^{\rm M}_0$ is a
function of $s$ and $\rho$, with the temperature $T$
and the chemical potential $\xi_0$ defined
respectively as $T\equiv\partial u^{\rm M}_0/\partial
s$ and $\xi_0\equiv\partial u^{\rm
M}_0/\partial\rho$. (The dimension of the chemical
potential $\xi_0$ is energy/mass.) The explicit form
of $u^{\rm M}_0$ is usually unknown, it depends on
microscopic specifics, and is in general rather
complicated. Yet, as we know from thermodynamics, and
as we shall see below, Eq~(\ref{21}) is as such
already very useful.

The second of Eqs~(\ref{5}),
$\mbox{\boldmath$g$}^{\rm M}
=\rho\mbox{\boldmath$v$}$, remains an excellent
approximation, because the energy flux ${\bf Q}^{\rm
M}$ as given by Eq~(\ref{add1}) is still dominated by
the term $\rho c^2\mbox{\boldmath$v$}$ (contained in
$u^{\rm M}{\boldsymbol v}$), from the transport of
rest energy. All the other terms are (in comparison)
relativistically small and can be neglected.

In nonrelativistic physics, it is customary to subtract
${\rm d}(c^2\rho)=c^2{\rm d}\rho$ from both sides of Eq
(\ref{21}), eliminating the rest energy from ${\rm
d}u^{\rm M}_0$ on the left, redefining the chemical
potential as $\xi_0-c^2$ on the right, and at the same
time subtracting $c^2\rho \mbox{\boldmath$v$}$ from the
energy flux ${\bf Q}^{\rm M}$, Eq~(\ref{add1}). This
represents a shift to a different but equivalent set of
independently conserved quantities: from $u^{\rm M}$ and
$\rho$ to $(u^{\rm M}-\rho c^2)$ and $\rho$. However,
this operation does alter the link between the momentum
density and the energy flux, rendering it
$\mbox{\boldmath$g$}^{\rm M}= {\bf Q}^{\rm
M}/c^2+\rho\mbox{\boldmath$v$}$ $\approx\rho {\boldsymbol
v}$. We shall follow this convention, so $u^{\rm M}_0$
and $\xi_0$, starting from this section, no longer
include the rest energy and $c^2$, respectively.

To account for a liquid with a varying local velocity
$v$, we need to employ a globally valid ``lab"-frame.
The material energy is then given as $u^{\rm
M}=u^{\rm M}_0+\rho v^2/2$, where ${\rm
d}(\textstyle\frac{1}{2}\rho v^2)=$ ${\rm
d}({\boldsymbol g}^{\rm M}/2\rho)=$
$\mbox{\boldmath$v$} \cdot{\rm
d}\mbox{\boldmath$g$}^{\rm M}-
\textstyle\frac{1}{2}v^2{\rm d}\rho$. So the
thermodynamic expression of Eq~(\ref{21}) becomes $\
{\rm d}u^{\rm M} = T{\rm d}s+\xi {\rm d} \rho
+\mbox{\boldmath$v$}\cdot {\rm
d}\mbox{\boldmath$g$}^{\rm M}\ $, with $\xi=$ $\xi_0-
\frac{1}{2}v^2$ the chemical potential of the lab
frame. Clearly, this expression signals that $u^{\rm
M}=u^{\rm M}(s,\rho_\alpha,\mbox{\boldmath$g$}^{\rm
M})$ is also a function of $\mbox{\boldmath$g$}^{\rm
M}$.

If we are dealing with a solution, a suspension, or
any system with more than one conserved densities, we
need to replace the term $\xi{\rm d}\rho$ with
$\xi_\alpha{\rm d}\rho_\alpha$, introducing all
conserved densities $\rho_\alpha$ as independent
thermodynamic variables,
\begin{equation}\label{23}
{\rm d}u^{\rm M} = T{\rm d}s+\xi_\alpha {\rm d}
\rho_\alpha +\mbox{\boldmath$v$}\cdot {\rm
d}\mbox{\boldmath$g$}^{\rm M},
\end{equation}
where summation over $\alpha=$ 1,~2,~$\cdots$ is implied.
Being conserved, all $\rho_\alpha$ obey continuity
equations, so the first of Eqs~(\ref{26}) is to be
replaced by $\dot\rho_\alpha+$ ${\boldsymbol\nabla}\!
\cdot(\rho_\alpha{\boldsymbol v})=0$. Usually, it is more
convenient to retain the total density $\rho$ as a
variable. For a binary mixture with the solute density
$\rho_1$ and the solvent density $\rho_2$, we therefore
take $\rho_1$ and $\rho\equiv\rho_1+\rho_2$ as the
variables, implying that $\xi_\alpha{\rm d}\rho_\alpha$
in Eq~(\ref{23}) is to be read as $\xi{\rm d}\rho$
$+\xi_1{\rm d}\rho_1$.

The pressure, defined as the energy change if the
volume changes at constant entropy and mass, is
related to the variables and conjugate variables of
$u^{\rm M}$ via the Duhem-Gibbs (or Euler) relation,
\begin{eqnarray}\label{31}
P\equiv-\partial({\textstyle\int} u_0^{\rm M}{\rm
d}^3r)/\partial{V} = -u^{\rm M}_0+Ts+(\xi_0)_\alpha
\rho_\alpha \nonumber\\ =-u^{\rm M}+Ts+\xi_\alpha
\rho_\alpha +\mbox{\boldmath$v$}
\cdot\mbox{\boldmath$g$}^{\rm M}.\quad
\end{eqnarray}
Combining Eq~(\ref{23}) with (\ref{31}) yields
\begin{equation}\label{31b}
{\boldsymbol\nabla} P=s{\boldsymbol\nabla} T+\rho_\alpha
{\boldsymbol\nabla}\xi_\alpha +g^{\rm
M}_j{\boldsymbol\nabla} v_j.
\end{equation}
Inserting this and the first of Eq~(\ref{26}) in
(\ref{28}), we find
\begin{eqnarray} \rho\textstyle\frac{\rm d}{{\rm
d}t}\mbox{\boldmath$v$} =\rho_e ({\bf e}+
\mbox{\boldmath$v$}\times{\bf b}/c)\nonumber\\
-(s{\boldsymbol\nabla} T+\rho_\alpha
{\boldsymbol\nabla}\xi_\alpha +g^{\rm
M}_j{\boldsymbol\nabla} v_j), \label{31a}
 \end{eqnarray}
an equation that will prove useful later. Note an
interesting feature of the hydrodynamic equations of
neutral systems, Eqs~(\ref{26},\ref{27},\ref{31a})
with $\rho_e\to0$: They are given in terms only of
the quantities appearing in Eq~(\ref{23}): the energy
$u^{\rm M}$, the thermodynamic variables
$s,\rho_\alpha, \mbox{\boldmath$g$}^{\rm M}$, and the
conjugate variables
$T,\xi_\alpha,\mbox{\boldmath$v$}$. Without an
explicit expression for $u^{\rm M}$, the hydrodynamic
theory is clearly necessarily written in these
general and abstract quantities, and we may take this
observation as an indication that the hydrodynamic
theory contains no more input than conservation laws
and thermodynamics, which is the basic reason of its
general validity. We shall return to, and build upon,
this point in the next section.

For a charged system, $\rho_e\not=0$, we need to include
the Maxwell equations (\ref{1},\ref{1a}) for a complete
description. The independent variables now include ${\bf
e}$ and $\bf b$, each with an equation of motion. All
equations also depend on $\rho_e$, which however only
stands for $\bf{\boldsymbol\nabla}\cdot e$ and is not
independent.

\subsubsection{``2-field" Theory}\label{hydr}
For our purpose of deriving the genuine low-resolution
Maxwell theory, it is instructive to again change the
notation to $u^{\rm tot}$ and $\mbox{\boldmath$g$}^{\rm
tot}$ by adding up the material and field contributions,
as we did in section~\ref{cons}. The new theory remains
closed and complete, and is given by the Maxwell
equations~(\ref{1},\ref{1a}), the continuity
equations~(\ref{26}) for mass and entropy, and
Eqs~(\ref{16}) for  $u^{\rm tot}$ and
$\mbox{\boldmath$g$}^{\rm tot}$, where the fluxes
\begin{eqnarray}\label{36}
{\bf Q}^{\rm tot}&=&(Ts+\xi_\alpha \rho_\alpha
+\mbox{\boldmath$v$} \cdot\mbox{\boldmath$g$}^{\rm M})
\mbox{\boldmath$v$} + c\,{\bf e}\times{\bf b},\\
\Pi_{ik}^{\rm tot}&=&(-u^{\rm tot}+ Ts+ \xi_\alpha
\rho_\alpha+
\mbox{\boldmath$v$}\cdot\mbox{\boldmath$g$}^{\rm M}+{\bf
e}^2+{\bf b}^2) \delta_{ik}\nonumber\\ &&+g^{\rm
M}_iv_k-e_ie_k-b_ib_k \label{37}
\end{eqnarray}
are obtained by adding the fluxes of Eq~(\ref{12}) and
(\ref{15}) to that of (\ref{add1}). The total momentum
density is still as in Eq~(\ref{18a}),
$\mbox{\boldmath$g$}^{\rm tot} =\rho\mbox{\boldmath$v$}+
{\bf e\times b}/c$, while the energy ${\rm d}u^{\rm tot}$
is obtained by adding Eq~(\ref{23}) to the differential
of the first of Eqs~(\ref{11}), yielding
\begin{eqnarray}\label{32}
{\rm d}u^{\rm tot}=T{\rm d}s+\xi_\alpha {\rm
d}\rho_\alpha + \mbox{\boldmath$v$}\cdot {\rm
d}\mbox{\boldmath$g$}^{\rm M} +{\bf e}\cdot{\rm d}{\bf e}
+{\bf b}\cdot{\rm d}{\bf b},\quad\\ =T{\rm
d}s+\xi_\alpha{\rm d}\rho_\alpha
+\mbox{\boldmath$v$}\cdot {\rm d}\mbox{\boldmath$g$}^{\rm
tot} +{\bf e}_0\cdot{\rm d}{\bf e} +{\bf b}_0\cdot{\rm
d}{\bf b}.\quad \label{33}\end{eqnarray} Eq~(\ref{33}) is
algebraically identical to Eq~(\ref{32}), because
$\mbox{\boldmath$v$}\cdot{\rm d}(\mbox{\boldmath$g$}^{\rm
M}-\mbox{\boldmath$g$}^{\rm tot})c=
-\mbox{\boldmath$v$}\cdot{\rm d}({\bf e}\times{\bf b}) =
-(\mbox{\boldmath$v$} \times{\bf e}) \cdot{\rm d}{\bf b}
+(\mbox{\boldmath$v$}\times{\bf b})\cdot{\rm d}{\bf e}$,
where ${\bf e}_0\equiv{\bf e}
+(\mbox{\boldmath$v$}/c)\times {\bf b}$,  ${\bf
b}_0\equiv{\bf b} -(\mbox{\boldmath$v$}/c) \times {\bf
e}$ are the respective rest frame fields. For reasons
that will become clear soon, we shall refer to this set
of equations as the ``2-field" theory.

In the low-resolution theory of the next chapter, we
shall be dealing with a dense system containing
hidden charges and dipole moments. This entails
coarse-graining operations, of which the most obvious
result is the appearance of four fields, $\bf E, D,
B, H$, replacing the two here. For the remaining of
this chapter, however, we shall ignore this
complication, while considering the less obvious, but
not less important changes.

Of these, the most relevant one is that we no longer
have an explicit expression for the field energy
$u^{\rm F}$; moreover, it is no longer possible to
distinguish $u^{\rm F}$ from $u^{\rm M}$, such that
the former only depends on the field variables and
the latter only on the hydrodynamic variables. (We
may of course expand $u^{\rm F}$ in the low-field
limit, and retain only the lowest order term, say
$\frac{1}{2}B^2/\mu$. Although this resembles the
microscopic expression, $\frac{1}{2}b^2$, the
coefficient $\mu$ is a function of hydrodynamic
variables.) Only $u^{\rm tot}=$ $u^{\rm F}+u^{\rm M}$
retains a clear-cut significance, because it is
conserved. This is relevant, as the form of $u^{\rm
F}$ was an important input of the electromagnetic
part of our theory, cf the last paragraph of
section~\ref{fiel}. On the other hand, this is not an
entirely new circumstance, as we did also loose the
knowledge about the explicit form of $u^{\rm M}$ when
going over to the dense, second system, cf
Eq~(\ref{21}).

So we shall proceed in two steps. First, we
reconsider the hydrodynamic theory of a neutral
system to understand its derivation, especially how
the ignorance about the energy was overcome. Next,
this method is generalized to include the $\bf e, \bf
b$-fields. The result will be a low-resolution
Maxwell theory --- except that two instead of four
fields are considered, or the 2-field theory.

In equilibrium, the entropy $s$ is a function of all
the conserved quantities in the system, because these
are the only ones that do not change when the system
is closed. In our case, the variables are $u^{\rm M},
g_i^{\rm M}, \rho_\alpha$. Given some (usually
satisfied) mathematical properties of $s(u^{\rm M},
g_i^{\rm M}, \rho_\alpha)$, we may rewrite this as
$u^{\rm M}(s, \rho, g_i^{\rm M})$, or Eq~(\ref{23}).
[Note that, at this stage, we do not yet know the
relation between $g_i^{\rm M}$ and
$\mbox{\boldmath$v$}\equiv\partial u^{\rm M}/\partial
g_i^{\rm M}$.]

As remarked in the last section, the hydrodynamic
equations are given in terms of the thermodynamic
variables and the associated conjugate variables.
This is closely related to the algebraic identity
given by the time derivative of Eq~(\ref{23}),
\begin{equation}\label{ffd}
\dot u^{\rm M}=T\dot s+\xi_\alpha\dot\rho_\alpha +
v_i\dot g_i^{\rm M},
\end{equation}
which the equations of
motion~(\ref{26},\ref{27},\ref{28}) must satisfy,
\begin{equation}\label{ffe}
{\boldsymbol\nabla}\cdot{\bf Q}^{\rm M}=T
{\boldsymbol\nabla}\cdot(s\mbox{\boldmath$v$})
+\xi_\alpha{\boldsymbol\nabla}\cdot
(\rho_\alpha\mbox{\boldmath$v$}) +v_i\nabla_j\Pi^{\rm
M}_{ij}.
\end{equation}
This is a very confining expression, and although not all
that obvious, does uniquely determine the fluxes ${\bf
Q}^{\rm M}$ and $\Pi^{\rm M}_{ij}$ with the help of some
auxiliary thermodynamic considerations. (This is
frequently referred to as the hydrodynamic standard
procedure.) So the hydrodynamics is indeed determined
once the thermodynamics is; there is an one-to-one
correspondence between them.

Given the explicit form of ${\bf Q}^{\rm M}$, the
equality $\mbox{\boldmath$g$}^{\rm M} =\rho
\mbox{\boldmath$v$}$ may now be established via
$\mbox{\boldmath$g$}^{\rm M}={\bf Q}^{\rm M}/c^2$.
Had ${\bf Q}^{\rm M}$ been different for some
reasons, so would $\mbox{\boldmath$g$}^{\rm M}$, but
the hydrodynamic theory -- in terms of $g_i^{\rm M}$
and $\mbox{\boldmath$v$}\equiv\partial u^{\rm
M}/\partial g_i^{\rm M}$ independently -- would
remain formally unchanged.

Including charges and fields, the conserved
quantities are $u^{\rm tot}, \rho_\alpha$ and
$\mbox{\boldmath$g$}^{\rm tot}$. So the entropy will
depend on them and the two fields, $s(u^{\rm tot},
\rho_\alpha, \mbox{\boldmath$g$}^{\rm tot}, {\bf e,
b})$. Again, this is equivalent to $u^{\rm tot}(s,
\rho_\alpha, \mbox{\boldmath$g$}^{\rm tot}, {\bf e,
b})$ -- compare this to Eq~(\ref{33}) to find that we
have already made the right choice of variables for
the ``2-filed" theory. What is more, given the
cogent, one-to-one relationship between the
thermodynamics and hydrodynamics, the associated set
of differential equations also possesses the right
structure of the low-resolution theory. This is
fortunate, as all there is left to do is to find out
how the two fields turn into four. That is, we may
now construct the genuine low-resolution theory by
requiring it to reduce to the present one in the
2-field limit, ${\bf E},\,{\bf D}\to{\bf e}$ and
${\bf H},\,{\bf B}\to{\bf b}$, (implying also ${\bf
E_0},\,{\bf D_0}\to{\bf e_0}$ and ${\bf H_0},\,{\bf
B_0}\to{\bf b_0}$).

One point remains to be settled: Given the algebraic
identity,
\begin{equation}\label{fff}
\dot u^{\rm tot}=T\dot s+\xi_\alpha\dot\rho_\alpha+
v_i\dot g_i^{\rm tot}+{\bf e_0 \cdot\dot e}+{\bf b_0
\cdot\dot b},
\end{equation}
the ``2-field" theory must be given only in terms of
the quantities in this equation. Unfortunately, the
two fluxes, Eqs~(\ref{36}, \ref{37}), do not seem to
conform to this rule and contain
$\mbox{\boldmath$g$}^{\rm M}$. This is alright at
present, because $\mbox{\boldmath$g$}^{\rm
M}=\rho\mbox{\boldmath$v$}$ is given in terms of two
quantities that do appear in Eq~(\ref{fff}), but it
seems to spell disaster for the low-resolution
theory, because $\mbox{\boldmath$g$}^{\rm M}$ is
without a clear meaning there.

Similar to the case for $u^{\rm tot}$, there is no
unambiguous division of $\mbox{\boldmath$g$}^{\rm
tot}$ into $\mbox{\boldmath$g$}^{\rm M}$ and
$\mbox{\boldmath$g$}^{\rm F}$ in systems with hidden
charges (although it is being attempted over and over
again with predictably rather contradictory results).
This is because $\mbox{\boldmath$g$}^{\rm M}$ lacks
any physical significance: First,
$\mbox{\boldmath$g$}^{\rm M}$ is devoid of its
significance as the conserved material momentum
density of a neutral system. Second, without a
uniquely defined $u^{\rm M}$ and the associated
energy current ${\bf Q}^{\rm M}$, neither is
$\mbox{\boldmath$g$}^{\rm M}={\bf Q}^{\rm M}/c^2$
useful as a work-around. (Not to mention that this
equality is invalid, since
$\mbox{\boldmath$\ell$}^{\rm M}$ is no longer
conserved). The same holds for
$\mbox{\boldmath$g$}^{\rm F}$. Nevertheless, one
clear-cut definition of $\mbox{\boldmath$g$}^{\rm M}$
does survive: Because of the algebraic equivalence
between Eqs~(\ref{32}) and (\ref{33}), we may start
from the second expression to obtain the first, {\em
defining} $\mbox{\boldmath$g$}^{\rm M}\equiv
\mbox{\boldmath$g$}^{\rm tot} - {\bf e\times b}/c$ as
a shorthand. Clearly, this guarantee that the two
fluxes, Eqs~(\ref{36}, \ref{37}), are still given by
the quantities of Eq~(\ref{fff}), even if
$\mbox{\boldmath$g$}^{\rm M}$ is no longer given by
$\rho\mbox{\boldmath$v$}$.

Finally, a few remarks about the pressure. While $T$ and
$\xi$, as indicated by Eq~(\ref{33}), remain well defined
as conjugate variables, and hence retain their place in
the low-resolution theory of the next section, the
pressure does not. [This is the reason we have chosen to
eliminate it from Eqs~(\ref{31a},\ref{37}).] One might
think of taking $P\equiv-\partial\int u_0^{\rm tot}{\rm
d}^3r/\partial{V}$, instead of Eq~(\ref{31}), only this
derivative is ambiguous and ill-defined: A dense system
in the presence of field is anisotropic, so the energy
not only depends on the volume, as implied by this
definition, but also on the shape. And the appropriate
quantity to deal with here is not the pressure, but the
stress tensor.

In isotropic liquids,  the pressure encompasses many
concepts that we find convenient, even intuitive: as
the surface force density, as the momentum current,
as a quantity that is continuous across interfaces,
and as a function of $T$ and $\rho$. Hence there is
widespread reluctance to abandon the pressure at
finite fields. Unfortunately, though there are
numerous ways to generalize the pressure that will
preserve some of these properties, none covers all.
So one may either define many different pressures --
an approach we eschew as it requires exceeding care
and has led to considerable confusion in the
literature -- or face up to the Maxwell stress
tensor, as we shall do here.

\subsection{Hydrodynamic Maxwell Theory}\label{EoM}

The theory we are going to consider, and the
expressions obtained hereby, are fairly general: They
are valid for arbitrarily strong fields and nonlinear
constitutive relations; the medium may be moving, and
the electromagnetic field may depend on time. But we
do assume the validity of local equilibrium, which
needs the characteristic time $\tau$ to be
established. Therefore, all frequencies are confined
to $\omega\tau\ll1$. (We emphasize that the theory is
only complete after the dissipative terms have been
included. Although the terms given here retain their
form off global equilibrium, at finite frequencies,
we need to exercise caution when applying them
alone.)

To derive the general theory, we need, as in the previous
two cases, the input of the Maxwell equations and some
information on the energy content of the field. The
macroscopic Maxwell equations are,
\begin{eqnarray}\label{48}
{\bf{\boldsymbol\nabla} \cdot D}= \rho_e,
\qquad\qquad\quad{\bf{\boldsymbol\nabla} \cdot B}= 0,  \\
\label{49}{\bf\dot D}=c{\boldsymbol\nabla}\times{\bf
H}-\rho_e\mbox{\boldmath$v$},\quad{\bf \dot
B}=-c{\boldsymbol\nabla}\times{\bf  E}.
\end{eqnarray}
Taking $D, B$ as the  variables, the first two are
constraints, and the next two their equations of motion.
If $\bf E$ and $\bf H$ are not specified, Eqs~(\ref{49})
do not contain any information beyond charge
conservation, as they may actually be derived from
$\dot\rho_e+{\boldsymbol\nabla}\cdot (\rho_e
\mbox{\boldmath$v$}) =0$ and Eqs~(\ref{48}). Especially,
no restriction whatever is implied for the two ``fluxes"
$\bf E$ and $\bf H$.

Taking the temporal derivative of the second constraint
yields ${\boldsymbol\nabla}\cdot{\bf\dot B}=0$, implying
there is a vector field $c{\bf E}$ such that the
divergence-free field ${\bf\dot B}$ may be written as
${\bf\dot B}= {\boldsymbol\nabla}\times (-c{\bf  E})$.
Similarly, deriving the first constraint yields
${\boldsymbol\nabla}\cdot({\bf\dot D}+\rho_e
\mbox{\boldmath$v$}) =0$, so this divergence-free field
must relate to a field $c{\bf H}$, such that ${\bf\dot
D}+\rho_e\mbox{\boldmath$v$}= {\boldsymbol\nabla} \times
(c{\bf H})$.

The temporal Maxwell equations~(\ref{49}) are given
content only by taking $\bf E$ and $\bf H$ as
thermodynamic conjugate variables, $\partial u^{\rm
tot}/\partial{\bf D}$ and $\partial u^{\rm
tot}/\partial{\bf B}$, respectively. More precisely,
we assert that, in the rest frame, the thermodynamic
energy has the form
\begin{equation}\label{rfe1}
{\rm d}u^{\rm tot}_0=T{\rm d}s+(\xi_0)_\alpha{\rm
d}\rho_\alpha+ {\bf E}_0\!\cdot{\rm d}{\bf D}_0 +{\bf
H}_0\!\cdot{\rm d}{\bf B}_0.
\end{equation}
It should not come as a surprise that we would need
this as an input, since the field dependence of
$u^{\rm F}$ was also an input, cf the last paragraph
of section~\ref{fiel}. (Some textbooks seem able to
``derive" the expression ${\bf E}\cdot{\rm d}{\bf D}
+{\bf H}\cdot{\rm d}{\bf B}$ from the Maxwell
equations, but this evidently contradicts our simple
consideration above.) The Poynting vector is obtained
by inserting the equations of motion (\ref{16},
\ref{26}, \ref{49}), for $\mbox{\boldmath$v$}=0$,
into Eq~(\ref{rfe1}), resulting in $\dot u_0^{\rm
tot}=-\nabla\cdot{\bf Q}_0^{\rm tot}$, with ${\bf
Q}_0^{\rm tot} =c {\bf E}_0\times{\bf H_0}$.

Knowing the rest frame quantities, it is not difficult in
principle to deduce the corresponding ones for the lab
frame of a moving liquid, $v\not=0$, although this does
entail some algebra. Fortunately, having established the
2-field limit, we have prepared ourselves a shortcut.
(The results of the dilute systems of course display the
correct transformation behavior.) First,
Eqs~(\ref{33},\ref{rfe1}) leave little choice other than
\begin{eqnarray}
{\rm d}u^{\rm tot}=T{\rm d}s+\xi_\alpha{\rm
d}\rho_\alpha+ \mbox{\boldmath$v$}\cdot {\rm d}
\mbox{\boldmath$g$}^{\rm tot}\nonumber\\ +{\bf
E}_0\cdot{\rm d}{\bf D} +{\bf H}_0\cdot{\rm d}{\bf
B}.\label{40}
\end{eqnarray}
[The algebraically minded reader may prefer to deduce
this expression directly. The appendix in
section~\ref{app} is specifically written for him.]
Second, since the rest frame fields are ${\bf E}_0=$
${\bf E} +(\mbox{\boldmath$v$}/c)\times {\bf B}$ and
${\bf H}_0 ={\bf H} -(\mbox{\boldmath$v$}/c)\times
{\bf D}$, we may rewrite Eq~(\ref{40}) as
\begin{eqnarray}
\label{42}{\rm d}u^{\rm tot} =T{\rm d}s+\xi_\alpha{\rm
d}\rho_\alpha+\mbox{\boldmath$v$}\!\cdot\!{\rm
d}\mbox{\boldmath $g$}^{\rm M} +{\bf E}\!\cdot\!{\rm
d}{\bf D} +{\bf H}\!\cdot\!{\rm d}{\bf B},\quad\\
\label{43}\qquad\mbox{\boldmath$g$}^{\rm
M}\equiv\mbox{\boldmath$g$}^{\rm tot}- {\bf D}\times{\bf
B}/c,\quad\quad
\end{eqnarray}
Eq~(\ref{42}) is the generalization  of
Eq~(\ref{32}), with $\mbox{\boldmath$g$}^{\rm M}$ now
given by Eq~(\ref{43}). As discussed, this does not
imply a field contribution of ${\bf D}\times{\bf
B}/c$, though it does tells us that this new
$\mbox{\boldmath$g$}^{\rm M}$ replaces the old one in
the fluxes of Eqs~(\ref{39},\ref{51x}) below. Note
that all conjugate variables are now functions of all
thermodynamic variables, eg $T$ is a function also of
${\bf D, B}$, and conversely, ${\bf H}_0$ depends on
$s,\rho$.

The rest frame Poynting vector $c{\bf E}_0\times{\bf
H_0}$ and Eq~(\ref{36}) makes it unambiguous that the
energy flux, and hence the momentum density, now have the
form,
\begin{eqnarray}\label{39}
{\bf Q}^{\rm
tot}=&&(Ts+\xi_\alpha\rho_\alpha+\mbox{\boldmath$v$}
\cdot\mbox{\boldmath$g$}^{\rm M})\mbox{\boldmath$v$} +
c\,{\bf E}\times{\bf H},\\
\label{50}&&\mbox{\boldmath$g$}^{\rm tot} =
\rho\mbox{\boldmath$v$}+{\bf E}\times{\bf H}/c,
\end{eqnarray} where
$[Ts+\xi_\alpha\rho_\alpha+\mbox{\boldmath$v$}
\cdot\mbox{\boldmath$g$}^{\rm M}]/\rho c^2\ll1$ is again
neglected in $\mbox{\boldmath$g$}^{\rm tot}$. (Being a
term of zeroth order in the velocity, ${\bf E}\times{\bf
H}/c$ may not be neglected with the same argument.)

To derive the stress tensor, the  last unknown expression
of the low resolution theory, we again take the temporal
derivative of Eq~(\ref{40}), $\dot u^{\rm tot} =T\dot s
+\xi_\alpha\dot \rho_\alpha+\cdots$ and insert the
respective equations of motion, Eqs~(\ref{16},\ref{26},
\ref{49}) and Eq~(\ref{39}), to arrive at
\begin{eqnarray}\label{39b}
{\boldsymbol\nabla}\cdot
[(Ts+\xi_\alpha\rho_\alpha+\mbox{\boldmath$v$}
\cdot\mbox{\boldmath$g$}^{\rm M})
 \mbox{\boldmath$v$} + c\,{\bf E} \times{\bf H}]\\
=T{\boldsymbol\nabla}\cdot(s\mbox{\boldmath$v$})+\xi_\alpha
{\boldsymbol\nabla} \cdot(\rho_\alpha
\mbox{\boldmath$v$}) +v_i\nabla_k\,\Pi_{ik}^{\rm
tot}\nonumber\\ -{\bf
E}_0\cdot(c{\boldsymbol\nabla}\times{\bf
H}-\rho_e\mbox{\boldmath$v$}) +c{\bf
H}_0\cdot{\boldsymbol\nabla}\times{\bf  E}.\nonumber
\end{eqnarray}
This equation is satisfied by the stress tensor
\begin{eqnarray}
&&\Pi_{ik}^{\rm tot}=\Pi_{ki}^{\rm tot}=g^{\rm
M}_iv_k-E_iD_k-H_iB_k+\nonumber\\&&(-u^{\rm tot}+ Ts+
\xi_\alpha\rho_\alpha+
\mbox{\boldmath$v$}\cdot\mbox{\boldmath$g$}^{\rm M} +
{\bf E\cdot D}+ {\bf H\cdot B}) \delta_{ik}.\quad
\label{51x}\end{eqnarray} Although this calculation only
yields $\nabla_k\Pi_{ik}^{\rm tot}$, so one can always
add a term $\epsilon_{kmn}\nabla_mA_{in}$ (with $A_{in}$
arbitrary), the requirement that it must reduce to
Eq~(\ref{37}) in the 2-field limit eliminates this
ambiguity.

Much information is contained in Eq~(\ref{39b}), and one
could indeed have derived the complete hydrodynamics
directly from it , without the detour over the dilute
system~\cite{henjes}. The path chosen in this article
highlights the physics and minimizes the algebra. [To
understand how stiflingly restricting Eq~(\ref{39b}) is,
just try out a few terms not in $\Pi_{ik}^{\rm tot}$.]

Consider the rotational invariance of the energy $u^{\rm
tot}$ to see that $\Pi_{ik}$ is symmetric: Rotating the
system by an infinitesimal angle ${\rm
d}\mbox{\boldmath$\theta$}$, the scalars are invariant,
${\rm d}u^{\rm tot}, {\rm d}s, {\rm d}\rho_\alpha=0$,
while the vectors change according to ${\rm
d}\mbox{\boldmath $g$}^{\rm M} =\mbox{\boldmath $g$}^{\rm
M}\times{\rm d}\mbox{\boldmath$\theta$}$,  ${\rm d}{\bf
D}= {\bf D}\times{\rm d}\mbox{\boldmath$\theta$}$, ${\rm
d}{\bf B}= {\bf B}\times{\rm d}\mbox{\boldmath$\theta$}$.
Inserting these into Eq~(\ref{42}) yields,
\begin{equation}\label{64}
\mbox{\boldmath$v$}\times\mbox{\boldmath $g$}^{\rm M}
+{\bf E}\times{\bf D} +{\bf H} \times{\bf B}=0,
\end{equation}
where the left side is equal to $\epsilon_{ijk}
\Pi_{kj}^{\rm tot}$. (For vanishing fluid velocity
and linear constitutive relations,
$\mbox{\boldmath$v$}\equiv0$,  $\epsilon{\bf E}={\bf
D}$ and $\mu{\bf H}={\bf B}$, the symmetry of the
stress tensor is obvious, otherwise it is not.)

Aside from the dissipative terms that we have
consistently neglected, the expressions of this section
represent the complete and closed hydrodynamic theory of
dense systems that are charged or subject to external
fields. The independent variables are:
$s,\rho_\alpha,\mbox{\boldmath$g$}^{\rm tot},{\bf D, B}$,
with $u^{\rm tot}$ and $\mbox{\boldmath$g$}^{\rm tot}$
given by Eqs~(\ref{42}) and (\ref{50}), respectively. The
attendant equations of motion are the Maxwell equations
Eqs~(\ref{48},\ref{49}), and
\begin{eqnarray}
\dot\rho_\alpha+{\boldsymbol\nabla}\cdot(\rho_\alpha
\mbox{\boldmath$v$})=0, \quad \dot
s+{\boldsymbol\nabla}\cdot(s\mbox{\boldmath$v$})=0,
\label{26x}\\ \label{16x} \dot u^{\rm
tot}+{\boldsymbol\nabla}\cdot{\bf Q}^{\rm tot}=0,\quad
\dot g_i^{\rm tot}+\nabla_k\Pi_{ik}^{\rm tot}=0,
\end{eqnarray}
where ${\bf Q}^{\rm tot}$ and $\Pi_{ik}^{\rm tot}$
are given by Eqs~(\ref{39},\ref{51x}).

In comparison to the considerations in available
textbooks, the main advantage of the above derivation
is the proof of consistency of the equations of
motion with general principles: Especially
thermodynamics, transformation behavior, and
conservation laws. At the same time, the range of
validity of these equations are clarified. The
advantage does not lie so much in the additional,
frequency and velocity dependent terms, which are
frequently small -- although this will change once
dissipative terms are included.

\subsection{Boundary Conditions}\label{boun}
To solve the derived set of hydrodynamic equations,
boundary conditions are needed. In mathematics,
boun\-dary conditions represent information that is
extrinsic to the differential equations. They are
provided to choose one special solution from the manifold
of all functions satisfying the differential equations.
In physics, circumstances are frequently different. An
example are the Maxwell equations, for which the boundary
conditions are actually derived by integrating the
differential equations across the interface, yielding
\begin{equation}\label{53}
\triangle D_n,\ \triangle B_n,\ \triangle E_t,\ \triangle
H_t=0
\end{equation}
(provided surface charges and currents are absent). The
subscripts $_n$ and $_t$ denote the components normal or
tangential to the interface, and $\triangle A\equiv
A_L-A_R$ denotes the discontinuity of any quantity $A$,
from left to right of the interface, toward positive
spatial coordinates. Note that since the integration has
to take place in a frame in which the interface is
stationary, we have $v_n=0$ in all the boundary
conditions derived in this section.

The results of the integration are better termed
``connecting conditions'', since they connect the
behavior of the variables on both sides of the
interface. This is possible because the Maxwell
equations are valid also on the other side of the
boundary. In contrast, mathematical boundary
conditions make no supposition on the validity of the
differential equations outside the considered region
-- in a sense, they shield the region from outside
influence. An analogous integration of the energy and
momentum conservation, Eqs~(\ref{16x}), yields
\begin{equation}\label{54}
\triangle Q^{\rm tot}_n=0, \ \triangle\Pi_{tn}^{\rm
tot}=0, \ \triangle\Pi_{nn}^{\rm tot}=
\alpha(R_1^{-1}+R_2^{-1}).
\end{equation}
The  first two connecting conditions do not contain much
independent information: Assuming $\mbox{\boldmath$v$}=0$
for both sides of the interface, and inserting the
expressions for $Q_n, \Pi_{tn}$ from
Eqs~(\ref{39},\ref{51x}), we find them to be satisfied
automatically in equilibrium if the connecting conditions
of the fields are: Defining $n_i$ and $t_i$ as unit
vectors that are normal and tangential to the interface,
$\Pi_{tn}^{\rm tot}=\Pi_{ik}^{\rm tot}t_in_k$,
$H_t=H_it_i$, $E_t=E_it_i$, $B_n=B_kn_k$, $D_n=D_kn_k$,
and with $H_t$, $E_t$, $B_n$ and $D_n$ continuous, cf
Eqs~(\ref{53}), we have $\triangle Q^{\rm tot}_n=0$ and
\begin{equation}\label{54x}
\triangle\Pi_{tn}^{\rm tot}=-\triangle(H_tB_n+E_tD_n)=0.
\end{equation}
Being  a generalization of $\triangle
P=\alpha(R_1^{-1}+R_2^{-1})$, the third boundary
condition for $\Pi_{nn}^{\rm tot}=\Pi_{ik}^{\rm
tot}n_in_k$ is rather more useful and will be
employed below repeatedly. ($\alpha>0$ is the surface
tension, and $R_1,R_2$ the principle radii of
curvature. The term $\alpha(R_1^{-1}+R_2^{-1})$ is
the equilibrium surface current for the momentum
density, a singular sink or source for the bulk
current.)

This completes the derivation of the hydrodynamic theory.
Strictly speaking, this is all we need to predict the
behavior of polarizable systems, as these partial
differential equations will deliver unique solutions if a
sufficient number of appropriate initial and boundary
conditions are provided. This renders our initial
question concerning the force of electromagnetic origin,
although as yet unanswered, seemingly irrelevant.
However, physics does not comprise solely of mathematics,
and clear thinking about the valid expressions for the
electromagnetic force will give us considerable heuristic
power in prediction --- without having to solve the set
of partial differential equations each and every time.

\section{Stress, Force and Pressure}\label{gene2}
To find an expression for the force in the presence
of fields, let us first remind ourselves of its form
without fields, a quantity that revolves around the
pressure:
\begin{itemize}
\item Its gradient, ${\boldsymbol\nabla}P=s{\boldsymbol\nabla}
T+ \rho_\alpha {\boldsymbol\nabla}\xi_\alpha +g^{\rm
M}_j{\boldsymbol\nabla} v_j$, is the bulk force density,
the quantity that accelerates a neutral volume element in
Eq~(\ref{31a}).  ${\boldsymbol\nabla}P$ vanishes in
equilibrium if we neglect gravitation.
\item The pressure $P$ itself is a surface force density,
which contributes to force equilibrium for instance in
hydrostatics. $P$ remains finite in equilibrium.
\end{itemize}
Since the Maxwell stress $\Pi_{ik}^{\rm tot}$ is the
generalization of the pressure for polarizable media, we
expect its gradient to be a bulk force density, see
section ~\ref{force}, and a quantity that relates to the
stress itself to be a surface force density, see
section~\ref{surfF}. Presumably, as in the field free
case, the bulk force density vanishes in equilibrium if
gravitation is neglected, while the surface force density
remains finite.

Generally speaking, the magnetic ponderomotive force is
up to five orders of magnitude stronger than the electric
one. This is connected to the fact that their respective,
easily attainable values are similar in SI units: $\hat
E\approx10^7$V/m, $\hat H\approx10^7$A/m (ie
$E\approx30$, $H\approx10^4$, in $\sqrt{\rm J/m^3}$).
Also, both susceptibilities are similar in magnitude, and
do not usually exceed $10^4$, hence we have $\mu_0\hat
H^2\approx\epsilon_0\hat E^2\times10^5$. In addition to
the greater ease and safety of handling, this frequently
makes magnetic fields the preferred ones. In almost all
the formulas of this and the next chapter (except in
sections \ref{force} and \ref{ener}), the electric terms
are completely analogous to the magnetic ones, and the
former may be obtained from the latter simply by
employing the replacements
\begin{equation}\label{rp1}
\bf B\to D,\,H\to E,\,M\to P.
\end{equation}
For linear constitutive relations (abbreviated hereafter
as {\sc lcr}), they imply $\mu\to\epsilon$, $\chi^{\rm
m}\to\chi$. To render the formulas simple, we shall
therefore usually display only the magnetic terms.

\subsection{The Bulk Force Density}\label{force}

In this section, we shall calculate $\nabla_k\Pi^{\rm
tot}_{ik}$ and look for all terms in the momentum
conservation that alter the acceleration $\rho\frac{\rm
d}{{\rm d}t}\mbox{\boldmath$v$}$ of a volume element,
starting from the energy density Eq~(\ref{42}),
\begin{equation}\label{rfe}
{\rm d}u^{\rm tot} =T{\rm d}s+\xi_\alpha{\rm
d}\rho_\alpha+\mbox{\boldmath$v$}\!\cdot\!{\rm
d}\mbox{\boldmath $g$}^{\rm M} +{\bf E}\!\cdot\!{\rm
d}{\bf D} +{\bf H}\!\cdot\!{\rm d}{\bf
B}.\end{equation} As discussed in the last chapter,
the subscript $\alpha$ enumerates the conserved
densities of the system. If there are more than one,
a summation over $\alpha$ is implied: $\xi_\alpha{\rm
d}\rho_\alpha=\xi{\rm d}\rho$ for a one-component
fluid, and $\xi_\alpha{\rm d}\rho_\alpha=$ $\xi{\rm
d}\rho+$ $\xi_1{\rm d}\rho_1$ for a two-component one
such as ferrofluids, where $\rho_1$ is the density of
magnetic particles, $\rho_2$ that of the fluid, and
$\rho=$ $\rho_1+\rho_2$ the total density. (A
ferrofluid is a suspension of magnetic particles,
typically of 10 nm diameter. Note that $\rho_1$ is
the mass of particles over the total volume. Its
variation does not arise from compressing each
particle individually, but from increasing the number
of the particles in a given volume element. This
appears to be a recurrent misunderstanding in the
ferrofluid literature.) It is useful to introduce a
number of potentials,
\begin{eqnarray}\label{F}
&&F(T,\rho_\alpha,v_i,B_i,D_i)=u^{\rm tot}-T s-g_i^{\rm
M}v_i,
\\  \label{tildeF}
&&\tilde F(T,\rho_\alpha,v_i,H_i,E_i)=F-H_iB_i-E_iD_i,
\\ \label{G}&&G(T,\xi_\alpha,v_i,B_i,D_i)=F-\xi_\alpha\rho_\alpha,
\\  \label{tildeG}
&&\tilde G(T,\xi_\alpha,v_i,H_i,E_i) =\tilde
F-\xi_\alpha\rho_\alpha.
\end{eqnarray}
The gradient of $\tilde G$ is
\begin{eqnarray}\label{nablaG}
\nabla_k\tilde G=-s\nabla_k
T-\rho_\alpha\nabla_k\xi_\alpha -g_i^{\rm M}\nabla_k
v_i\\-B_i\nabla_k H_i-D_i\nabla_k E_i.\nonumber
\end{eqnarray}
Writing the Maxwell stress, Eq~(\ref{51x}), as
\begin{equation} \label{51}
\Pi_{ik}^{\rm tot}=-\tilde G\delta_{ik}+g_i^{\rm M}
v_k -H_iB_k-E_iD_k,\end{equation}
we find
\begin{eqnarray}\label{kraft2}
&&\nabla_k\Pi_{ik}^{\rm tot}=s\nabla_i
T+\rho_\alpha\nabla_i\xi_\alpha+B_k(\nabla_i H_k-
\nabla_k H_i)\\ &&+D_k(\nabla_i E_k- \nabla_k
E_i)-\rho_eE_i+g_k^{\rm M}\nabla_iv_k+\nabla_k(g_i^{\rm
M}v_k)\nonumber
\end{eqnarray}
which, in addition to the Maxwell equations (\ref{49}),
are to be inserted into the momentum conservation,
\begin{equation}\label{inclGrav}
\dot g_i^{\rm tot}+\nabla_k\Pi_{ik}^{\rm tot}=-\rho g{\bf
e_z}.
\end{equation}
[We shall from now on include gravitation, where $g$ is
the acceleration of gravity, and the unit vector ${\bf
e_z}$ points upward.] The result is
\begin{eqnarray}
\rho\textstyle\frac{\rm d}{{\rm
d}t}[\mbox{\boldmath$v$}+({\bf E\times H}- {\bf D\times
B})/c\rho] =\rho_e ({\bf E}+
\mbox{\boldmath$v$}\times{\bf
B}/c)\nonumber\\-(s{\boldsymbol\nabla} T+
 \rho_\alpha{\boldsymbol\nabla} \xi_\alpha
+g^{\rm M}_j{\boldsymbol\nabla} v_j)-\rho g{\bf
e_z}.\qquad
 \label{56}\end{eqnarray}
Compare this to Eq~(\ref{31a}) and register the great
similarity --- but do also remember that the temperature
$T$ and chemical potentials $\xi_\alpha$ are now
functions of the fields. All terms except
$\rho\textstyle\frac{\rm d}{{\rm d}t}\mbox{\boldmath$v$}$
may be taken as various force densities: The
frequency-dependent, so-called Abraham-force,
$-\rho\frac{\rm d}{{\rm d}t}({\bf E\times H}- {\bf
D\times B})/c\rho$, not usually a large term, the ``fake"
Lorentz force, $\rho_e ({\bf E}+
\mbox{\boldmath$v$}\times{\bf B}/c)$, only significant if
the system is macroscopically charged, $\rho_e\not=0$,
and the bulk force density
\begin{equation}\label{f}
{\bf f}^{\rm bulk}=-(s{\boldsymbol\nabla} T+\rho_\alpha
{\boldsymbol\nabla}\xi_\alpha+g^{\rm
M}_j{\boldsymbol\nabla} v_j)-\rho g{\bf e_z},
\end{equation}
which includes both the gravitational and the
electromagnetic force. Note ${\bf f}^{\rm bulk}=$
$-\nabla_k\Pi_{ik}^{\rm tot}$ $-\rho g{\bf e_z}$ in the
neutral, stationary limit, for ${\boldsymbol v}$,
${\boldsymbol\nabla}\times{\bf E}$,
${\boldsymbol\nabla}\times{\bf H}=0$, cf
Eq~(\ref{kraft2}). ${\bf f}^{\rm bulk}$ is the proper
macroscopic, coarse-grained force, valid as long as the
hydrodynamic, macroscopic Maxwell theory is. Together
with the Abraham force, it accounts for the difference
between the true and the ``fake" Lorentz force discussed
in the Introduction.  It consists only of
thermodynamically well defined quantities, either
variables or conjugate variables of Eq~(\ref{rfe}). This
is no longer the case if we follow a widespread
convention to write ${\bf f}^{\rm bulk}$ as a sum of the
zero-field pressure gradient and the ponderomotive (or
Kelvin) force, as this introduces thermodynamically
ill-defined quantities and unwelcome ambiguities, cf
section~\ref{pond}. In equilibrium, we have
\begin{equation}\label{Equil} {\boldsymbol\nabla}T,
{\boldsymbol\nabla}\xi_1=0,
\quad{\boldsymbol\nabla}\xi=-g{\bf e_z}.
\end{equation}
Inserting these in Eq~(\ref{f}), we find, for stationary
fluids in equilibrium, ${\bf f}^{\rm bulk}=0$.

The considerations of the last two chapters make
abundantly clear that the conserved momentum density is
the sum of material and field contribution, ${\boldsymbol
g}^{\rm tot}=$ $\rho{\boldsymbol v}+$ ${\bf E}\times{\bf
H}/c$, and that the corresponding flux is the Maxwell
tensor. Nevertheless, in the context of condensed matter,
because $\rho{\boldsymbol v}\gg$ ${\bf E}\times{\bf
H}/c$, the second term may usually be neglected --- and
since ${\bf D}\times{\bf B}$ and ${\bf E}\times{\bf H}$
are of the same order of magnitude, so is the Abraham
force. [Taking $\rho$ as 1 g/cm$^3$, $v$ as 1 cm/s, $H$
as $10^4$ and $E$ as 30, both in $\sqrt{\rm J/M^2}$, (ie
$\hat H=10^7$A/m, $\hat E=10^7$V/m,) we have $\rho vc/EH$
$\approx3000$.]

\subsection{The Incompressible Limit}\label{incomp}
It is noteworthy that the incompressible limit in
ferrofluids does not imply $\dot\rho$,
${\boldsymbol\nabla}\rho=0$, and hence
${\boldsymbol\nabla}\cdot {\boldsymbol v}=0$. This is
because incompressibility means the constancy of the two
local, actual densities, $\rho_{\rm M}$ of magnetic
particles and $\rho_{\rm F}$ of the fluid matrix (ie
$\rho_1=$ $\langle\rho_{\rm M}\rangle$ and $\rho_2=$
$\langle\rho_{\rm F}\rangle$, with the averaging taken
over a volume containing many particles). Yet since the
particles are usually denser than the fluid, $\rho_{\rm
M}\approx5\rho_{\rm F}$, an increase of particle
concentration will also increase the total density
$\rho=$ $\rho_1+\rho_2$. More quantitatively,
$\rho_1/\rho_{\rm M}+$ $\rho_2/\rho_{\rm F}$ $=1$ because
$\rho_1/\rho_{\rm M}$ is the fraction of volume occupied
by the particles, and $\rho_2/\rho_{\rm F}$ that occupied
by the fluid. Taking $\rho_{\rm M}$ and $\rho_{\rm F}$ as
constant in the incompressible limit, we have
\begin{equation}\label{inc}
{\rm d}\rho=\gamma{\rm d}\rho_1,\quad\gamma=1-\rho_{\rm
F}/\rho_{\rm M}.
\end{equation}
Inserting this expression into $\xi{\rm d}\rho+$
$\xi_1{\rm d}\rho_1$, we have $\xi^{\rm in}_1{\rm
d}\rho_1$ with $\xi^{\rm in}_1=$ $\xi_1+\gamma\xi$, and
the modified equilibrium conditions
\begin{equation}\label{equil}
{\boldsymbol\nabla} T=0,\quad{\boldsymbol\nabla}\xi^{\rm
in}_1=-g\gamma{\bf e_z}.
\end{equation}
Ferrofluids may frequently be approximated as being in
the incompressible limit considered here, because the
variations of $\rho_{\rm F}$ and $\rho_{\rm M}$, say as a
result of field inhomogeneities, are usually negligible.
The particle density $\rho_1$, on the other hand, may
vary greatly, and so does $\rho$ via Eq~(\ref{inc}). The
parameter controlling this behavior is the osmotic
compressibility, $\kappa_{\rm os}=
\rho_1^{-2}\partial\rho_1/
\partial\xi^{\rm in}_1$, which is larger by around 6 orders of
magnitude than the compressibility of an ordinary liquid,
cf section~\ref{denva}.

\subsection{The Maxwell Stress in Equilibrium}
\label{MSE} We consider the simplifications that
occur for the Maxwell stress tensor if the system is
stationary ($v\equiv0$) and in equilibrium. [Starting
from this section, we shall no longer display the
electric terms explicitly, ${\bf E,D}=0$. And the
subscript $_0$ denoting rest frame quantities will
also be eliminated.]

\subsubsection{Total Equilibrium}

With the equilibrium conditions as given in
Eqs~(\ref{Equil}), $\tilde G$ is nonuniform only due to
the inhomogeneities in the fields, both gravitational and
electromagnetic, cf Eq~(\ref{nablaG}). This we may
utilize to obtain a more handy expression for the stress.
Inserting Eq~(\ref{Equil}) in (\ref{nablaG}) and
integrating the resulting expression from 1 to 2, two
arbitrary points in the medium, we have
\begin{equation}\label{dB/dH}
\mbox{\boldmath$\Delta$}(\tilde G-\bar\rho
gz+{\textstyle\int_{\rm Eq}}B_i{\rm d}H_i)=0,
\end{equation}
where $\mbox{\boldmath$\Delta$}$ denotes the difference
between the two points of the quantity behind it, eg
${\boldsymbol\Delta}\tilde G\equiv\tilde G_2- \tilde
G_1$. The first in the above equation is from
${\boldsymbol\Delta} \tilde G=$ $\int_1^2\nabla_k\tilde
G{\rm d}r_k$; the magnetic term is from the integral
$\int^2_1(B_i\nabla_kH_i){\rm d}r_k$: Writing it as
$\int^2_1 B_i{\rm d}H_i$, we note that it is in fact a
difference of purely local quantities, with no reference
to the path: $\int^2_1 B_i{\rm d}H= \int^2_0 B_i{\rm
d}H_i-\int^1_0 B_i{\rm d}H_i\equiv \mbox{\boldmath
$\Delta$}\int B_i{\rm d}H_i$, where 0 denotes a (possibly
virtual) spot of vanishing field, $H_i=0$, and 1, 2
respectively the field values at the spatial points 1, 2.
The integration is unambiguous because it is to be
carried out in equilibrium, for constant $T,\xi_1$ and
$\xi+gz$, or for constant $T$, $\xi^{\rm in}_1+\gamma gz$
in the incompressible approximation. To emphasize this,
the subscript $\int_{\rm Eq}$ is added. With $M_i\equiv
B_i-H_i$, we may also write this term as
${\textstyle\frac{1}{2}H^2} +{\textstyle\int_{\rm
Eq}}M_i{\rm d}H_i$. The gravitational term comes from
integrating the chemical potential,
$-\int_1^2\rho\nabla_k\xi{\rm d}r_k=$ $g\int_1^2\rho{\rm
d}z=$ $g\int_0^2\rho{\rm d}z$ $ -g\int_0^1\rho{\rm d}z$
$=g{\boldsymbol\Delta} (z\bar\rho)$, where $z\bar\rho(z)$
$\equiv\int_0^z\rho{\rm d}z$.

Eq~(\ref{dB/dH}) states the constancy in equilibrium of
the quantity in the bracket, call it $-K$,
\begin{equation}\label{defK}
-K\equiv\tilde G- \bar\rho gz+ {\textstyle\frac{1}{2}H^2}
+{\textstyle\int_{\rm Eq}}M_i{\rm d}H_i.
\end{equation}
This enables us to write the stress tensor,
Eq~(\ref{51}), as
\begin{equation}\label{2dB/dH}
\Pi^{\rm tot}_{ij} = (K+{\textstyle\frac{1}{2}H^2}
+{\textstyle\int_{\rm Eq}}M_i{\rm d}H_i-\bar\rho gz)
\delta_{ij}-H_iB_j,
\end{equation}
the announced handy expression. As we shall see, it is
rather useful, and even includes the information
contained in the magnetic Bernoulli equation~\cite{rz},
see section~\ref{bernoulli}. More generally, we may
calculate $\Pi^{\rm tot}_{ij}$ for an arbitrary point
within the medium --- if we know the field everywhere and
the value of the constant K (usually via the boundary
conditions to be discussed below).

If the system under consideration (say the atmosphere) is
non-magnetic, $M_i=0$, $B_i=H_i$ and $u^{\rm tot}=u^{\rm
M}+\frac{1}{2}H^2$ hold. Inserting these into
Eq~(\ref{defK}) and employing Eqs~(\ref{31}), we have
\begin{equation}\label{K=P}
K=\bar\rho gz+P_{\rm atm},
\end{equation}
with $K$ clearly the pressure at $z=0$. If the system is
magnetic, we may still write $\tilde G(T,\xi_\alpha,H_i)=
\tilde G(0)+\tilde G_{\rm em}$, where the first denotes
$\tilde G$ for vanishing $H$, while the second the
contribution from the field,
\begin{equation}\label{mbe1}
-\tilde G_{\rm em}={\textstyle\int_{\rm Eq}}B_i{\rm
d}H_i={\textstyle\frac{1}{2}H^2}+{\textstyle\int_{\rm
Eq}}M_i{\rm d}H_i,
\end{equation}
and identify
\begin{equation}\label{mbe1a}
K=\bar\rho gz-\tilde G(0).
\end{equation}
For obvious reasons, $\tilde G(0)$ is frequently referred
to as the ``zero-field pressure". This is unfortunate, as
$\tilde G(0)$ $\equiv\tilde G$ $-\tilde G_{\rm em}$ is a
function of $T,\xi,\xi_1$ -- these are the natural
variables of $\tilde G$, and the value of the integral in
Eq~(\ref{mbe1}) explicitly depends on this choice of
variables. On the other hand, if the zero-field pressure
were a physically sensible quantity, it should remain the
same whichever variables it is taken to depend on. It
does not, of course, as it is equal to $\tilde G(0)$.
This renders the associated concept of pressure
ambiguous, ill defined, and best avoided, see
section~\ref{MN} for more details.

Combining Eq~(\ref{nablaG}) and (\ref{mbe1}) -- ignore
gravitation and the electric field -- we find
\begin{equation}\label{intForm}
{\boldsymbol\nabla}\int B_i{\rm d}H_i=
B_i{\boldsymbol\nabla}H_i.
\end{equation}
For a function of one variable, this relation always
holds: Defining $F=\int f(g) {\rm d}g$, we have
${\boldsymbol\nabla} F=$ $({\rm d}F/{\rm
d}g){\boldsymbol\nabla} g$ $=f{\boldsymbol\nabla}g$. For
a function that depends on more variables, this relation
holds only if the additional variables are spatially
constant. In the present case, $B$ does depend on
additional variables, $T$, $\xi_1$, $\xi+gz$.
Fortunately, all three are constant. [$B$ may also, as in
next section, depend on constant $T$, $\rho$ and $\rho_1$
--- which is the reason the subscript in $\int_{\rm Eq}$
is eliminated from Eq~(\ref{intForm}).] Considering {\sc
lcr}, we find ${\boldsymbol\nabla}{\textstyle\int}B_i{\rm
d}H_i=$ ${\boldsymbol\nabla}(\frac{1}{2}\mu H^2)$, which
is equal to $B_i{\boldsymbol\nabla}H_i$ if $\mu$ (say as
a function of $T$) is constant.

Turning our attention now to boundary conditions, we find
that the expression $\triangle\Pi^{\rm tot}_{nn}$ [which
occurs in the boundary condition Eqs~(\ref{54})] may be
written as
\begin{eqnarray}\label{301}
\triangle\Pi^{\rm tot}_{nn}=
\triangle(K+{\textstyle\int_{\rm Eq}}M_i{\rm
d}H_i+\textstyle\frac{1}{2}M_n^2-\bar\rho gz)
\end{eqnarray}
in equilibrium. Two remarks: (i) The equality
$\triangle(\textstyle\frac{1}{2}H^2-H_nB_n)
=\triangle(\textstyle\frac{1}{2}M_n^2
+\textstyle\frac{1}{2}H_t^2 -\textstyle\frac{1}{2}B_n^2)$
was used to derive Eq~(\ref{301}), where
$\triangle(H_t^2)$, $\triangle(B_n^2)$ vanish,
cf~Eq(\ref{53}). (ii) If appropriate, substitute
$\int_{\rm Eq}$ with $\int_{\rm uni}$, and $\bar\rho$
with $\rho$, cf Eq~(\ref{sn22}) below.

Inserting Eq~(\ref{301}) into the third of
Eqs~(\ref{54}), assuming atmosphere on one side,
employing Eq~(\ref{K=P}) while neglecting the
gravitational term $\bar\rho gz$ of the atmosphere, the
boundary condition reduces to
\begin{eqnarray}\nonumber
K+{\textstyle\int_{\rm Eq}}M_i{\rm
d}H_i+\textstyle\frac{1}{2}M_n^2-\bar\rho gz\\ =P_{\rm
atm}+\alpha(R_1^{-1}+R_2^{-1}),\label{K}
\end{eqnarray}
a useful formula that we shall frequently refer to below.

\subsubsection{Quasi-Equilibrium}
The considerations of the last section is not confined to
total equilibrium, and may be generalized to include
quasi-equilibria. Establishing equilibrium with respect
to the distribution of magnetic (or electric) particles
is a slow process, because mesoscopic particles diffuse
much more slowly than atomic ones. Depending on the field
gradient, the relevant volume and the size of the
particles, this may take days, even
weeks~\cite{blums,odenbach}. For a rough estimate, we
equate the Stokes with the Kelvin force to calculate the
velocity $v$ with which a magnetic particle moves:
$6\pi\eta Rv=$ $(4\pi R^3/3)\cdot$ $\chi\mu_0\nabla\hat
H^2/2$. Taking the particle radius as $R=10$nm, the
viscosity as $\eta=10^{-3}$kg/ms, the susceptibility as
$\chi\approx$1, the field as $\hat B=0.1$T, and the field
gradient $\nabla\hat B$ as 1T/mm, the velocity is around
$10^{-3}$mm/s, and the time the particles needs to
achieve equilibrium is $\tau\approx$ 1mm/$v$
$\approx10^3$s. On the other hand, particles $10^2$ times
larger, of the size of $1\mu$, lead to velocities $10^4$
times larger, and a characteristic time of $10^{-1}$s.
For time scales much smaller than $\tau$, after a uniform
ferrofluid is brought into contact with an inhomogeneous
magnetic field, the following conditions hold instead of
Eqs(\ref{Equil}),
\begin{equation}\label{sn}
{\boldsymbol\nabla}\rho_1=0,\, {\boldsymbol\nabla} T=0,\,
\rho_1{\boldsymbol\nabla}\xi_1+\rho{\boldsymbol\nabla}\xi=-\rho
g{\bf e_z}.
\end{equation}
This is because both the diffusion of heat and  the
establishment of mechanical equilibrium are in comparison
fast processes. The last of Eq~(\ref{sn}) states ${\bf
f}^{\rm bulk}=0$, cf Eq~(\ref{f}), which is due to the
density $\rho$ quickly turning slightly nonuniform to
compensate for the field inhomogeneity. Inserting
Eq~(\ref{sn}) in (\ref{nablaG}), we find
${\boldsymbol\nabla}\tilde G=$ $\rho g{\bf e_z}-$
$B_i{\boldsymbol\nabla} H_i$, an expression that we can
integrate directly, from a point 1 to a point 2, if we
approximate the variables $T$, $\rho$, $\rho_1$ as
constant, yielding
\begin{equation}\label{sn2}
{\boldsymbol\Delta}(\tilde G-\rho gz+
 {\textstyle\int_{\rm uni}}B_i{\rm d}H_i)=0,
\end{equation}
where the subscript $\int_{\rm uni}$ indicates that the
integral is evaluated at constant $T$, $\rho$, and
$\rho_1$. Again calling the spatially constant quantity
in the bracket $-K$, the stress tensor is
\begin{eqnarray}\label{sn22} \Pi^{\rm tot}_{ij} =
(K+{\textstyle\frac{1}{2}H^2} +{\textstyle\int_{\rm
uni}}M_i{\rm d}H_i-\rho gz) \delta_{ij}-H_iB_j,\quad\quad
\end{eqnarray}
where the subscript of the integral and the lacking bar
over $\rho$ are the only differences to
Eq~(\ref{2dB/dH}). Integrating as here the magnetization
while holding constant $T$ and $\rho_\alpha$ yields the
electromagnetic contribution to the free energy,
\begin{equation}\label{sn3}
-\tilde F_{\rm em}(T,\rho_\alpha)={\textstyle\int_{\rm
uni}}B_i{\rm d}H_i={\textstyle \frac{1}{2}H^2}
+{\textstyle\int_{\rm uni}}M_i{\rm d}H_i,
\end{equation}
so $K$ may be written as
\begin{equation}\label{mbe1b}
K=\rho gz-[\tilde F(0)-\xi_\alpha\rho_\alpha],
\end{equation}
while Eq~(\ref{K=P}) still holds for nonmagnetic media.

Note that $M_i(T,\rho_1,H_i)$ is the appropriate function
for evaluating $\int_{\rm uni}M_i{\rm d}H_i$, while
$M_i(T,\xi^{\rm in}_1,H_i)$ is the proper one for
$\int_{\rm Eq}M_i{\rm d}H_i$ -- both in the
incompressible approximation. The first is given by an
experi\-ment that quickly measures the magnetization
while varying the external field in a closed system, such
that $\rho_1$ stays constant -- though contact with a
heat bath is necessary to maintain the temperature. The
latter, $M_i(T,\xi^{\rm in}_1,H_i)$, is to be measured in
an open system that is connected to a particle reservoir,
which itself is not subject to a varying field, so its
chemical potentials $\xi^{\rm in}_1$ remains constant.
Increasing the field of the system now to measure the
magnetization, magnetic particles will enter the system
from the reservoir, resulting in a stronger magnetization
than in the closed case.

Either experiment is sufficient to determine both
magnetizations, as one can be calculated from the other.
This is accomplished by employing the thermodynamic
relation,
\begin{equation}\label{ti}
\left[\frac{\partial M}{\partial H}\right]_{\xi^{\rm
in}_1}= \left[\frac{\partial M}{\partial
H}\right]_{\rho_1} +\frac{\rho^2_1}{\kappa^{-1}_{\rm
os}}\left[\frac{\partial M}{\partial
\rho_1}\right]^2_{H},
\end{equation}
where the inverse osmotic pressure is given as
$\kappa_{\rm os}^{-1}=$ $\rho_1^{2}$ $\times$ $
(\partial\xi^{\rm in}_1/\partial\rho_1)_H$. Eq~(\ref{ti})
may be verified by combining a thermodynamic identity
with a Maxwell relation,
\begin{eqnarray}
\left[\frac{\partial M}{\partial H}\right]_{\xi^{\rm
in}_1}= \left[\frac{\partial M}{\partial
H}\right]_{\rho_1} +\left[\frac{\partial
M}{\partial\rho_1}\right]_{H}
\left[\frac{\partial\rho_1}{\partial H}\right]_{\xi^{\rm
in}_1},\\ \left[\frac{\partial\rho_1}{\partial
H}\right]_{\xi^{\rm in}_1} =\left[\frac{\partial
B}{\partial\xi^{\rm in}_1}\right]_{H}
=\left[\frac{\partial
M}{\partial\rho_1}\right]_{H}
\left[\frac{\partial\rho_1}{\partial\xi^{\rm
in}_1}\right]_{H}.
\end{eqnarray}
[We are assuming that $\bf M\| H$, so all $M$ and $H$
here are to be understood as the respective magnitude. In
addition, $T$ is held constant and $\rho$ is taken to
satisfy the incompressibility condition Eq~(\ref{inc})
throughout.] For {\sc lcr}, Eq~(\ref{ti}) reduces to
\begin{equation}\label{ti2}
\chi^{\rm m}(T,\xi^{\rm in}_1,\rho)=\chi^{\rm
m}(T,\rho_1,\rho)+\left(\rho_1\frac{\partial\chi^{\rm
m}}{\partial \rho_1}\right)^2\frac{H^2}{\kappa^{-1}_{\rm
os}},
\end{equation}
showing that the difference between the two
susceptibilities is of higher order in the field, and may
be neglected if one strictly adheres to  {\sc lcr}. On
the other hand, this calculation also shows when the
second term may no longer be neglected: Assuming for an
order of magnitude estimate that $\rho_1 (\partial
\chi^{\rm m}/\partial\rho_1)$ $\approx\chi^{\rm m}$
$\approx1$, and taking $\kappa_{\rm os}^{-1}$
$\approx10^{3}$~Pa (see the calculation in
section~\ref{denva} below), we find that a field of $\hat
H\approx300$~Oe suffices to render $H^2/\kappa^{-1}_{\rm
os}\approx1$.

\subsection{Equilibrium Surface Force}\label{surfF}

Given a solid, polarizable body, or one that is (although
not polarizable) submerged in ferrofluid, we may lift the
body off the ground electromagnetically, against the
gravitational force, or balance it against some elastic
force exerted by a string or spring. The same applies to
a vessel containing ferrofluid. All these happen even in
equilibrium, when the bulk force density, Eq~(\ref{f}),
vanishes. This force is
\begin{eqnarray}\label{emsf2}
{\pmb{\cal F}}^{\rm elm}=\triangle\oint [{\textstyle
\frac{1}{2}}M^2_n+ {\textstyle\int_{\rm Eq}}M_k{\rm
d}H_k]\,{\rm d}{\bf A}\\\label{emsf}
 =\triangle\oint[{\textstyle
  \int_{\rm Eq}}M_t{\rm d}H_t+{\textstyle
  \int_{\rm Eq}}M_n{\rm d}B_n]\,{\rm d} {\bf A},
\end{eqnarray}
with the electric counter terms as usual given by
Eq~(\ref{rp1}). $\triangle$ now specifically denotes:
internal - external, and the two integrals are to be
taken along the surface of the body, one right inside and
the other right outside of it. As in the last section,
${\textstyle\int_{\rm Eq}}$ is to be substituted by
${\textstyle\int_{\rm uni}}$, and $\bar\rho$ by $\rho$,
where appropriate.

Before deriving Eq~(\ref{emsf2}), we shall first consider
the ramifications of this formula in two examples: (i) A
magnetizable solid body (or a nonmagnetic vessel
containing ferrofluid) in atmosphere, (ii) a nonmagnetic
solid body in ferrofluid.

In the first example, the external magnetization is zero.
Employing the Gauss law, we change the surface integral
of Eq~(\ref{emsf}) into a volume integral: With
${\boldsymbol\nabla}{\textstyle\int_{\rm Eq}}M_t{\rm
d}H_t=$ $M_t{\boldsymbol\nabla}H_t$ and
${\boldsymbol\nabla}{\textstyle\int_{\rm Eq}}M_n{\rm
d}B_n=$ $M_n{\boldsymbol\nabla}B_n$, cf
Eq~(\ref{intForm}), the electromagnetic force is
\begin{equation}\label{emvf!}
{\pmb {\cal F}}^{\rm elm} =\int (M_t{\boldsymbol\nabla}
H_t+ M_n{\boldsymbol\nabla}B_n){\rm d}^3r.
\end{equation}
Consider a plate with the field gradient normal to its
surface: If the field is predominantly tangential to its
surface, the electromagnetic force is $\int
M_k{\boldsymbol\nabla} H_k{\rm d}^3r$; if the field is
normal to the plate's surface, it is $\int
M_k{\boldsymbol\nabla} B_k{\rm d}^3r$. So interestingly,
the magnetic force density interpolates between both
forms of the Kelvin force as discussed in the
introduction -- and in greater details in section
\ref{pond} below -- though with the difference that these
here are exact results, as no assumption was made with
respect to the constitutive relation, especially the
density dependence of the magnetization (or
susceptibility).

If the geometry is more complicated than in a plate, say
if the magnetizable body is an ellipsoid, it is less
clear what the normal and tangential component of $M$,
$H$ and $B$ are in the bulk, and Eq~(\ref{emvf!}) appears
ambiguous. Fortunately, this does not matter, because the
integral of Eq~(\ref{emvf!}) has (by virtue of the Gauss
law) a unique value, as long as $\int M_t{\rm d}H_t$ and
$\int M_n{\rm d}B_n$ are given at the surface, cf
Eq~(\ref{emsf}), and the fields $M_t$, $H_t$, $M_n$, and
$B_n$ being continuous in the bulk. [An analogous
situation is given by the simple integral $\int^2_1f{\rm
d}x$, $F'=f$, and $F(1)$, $F(2)$ fixed. Then
$\int^2_1f{\rm d}x$ $=F(2)$ $-F(1)$ is unique
irrespective of how $f$ varies in the interval between 1
and 2.]

Turning our attention now to the second system, a
non-polarizable body submerged in ferrofluid, we take the
internal magnetization as zero and find
\begin{eqnarray}\label{emsf3}
{\pmb{\cal F}}^{\rm elm}=-\oint [{\textstyle
\frac{1}{2}}M^2_n+ {\textstyle\int_{\rm Eq}}M_k{\rm
d}H_k]\,{\rm d}{\bf A}\\\label{emsf4}
 =-\oint[{\textstyle
  \int_{\rm Eq}}M_t{\rm d}H_t+{\textstyle
  \int_{\rm Eq}}M_n{\rm d}B_n]\,{\rm d} {\bf A}.
\end{eqnarray}
It may also here be useful to employ the Gauss law for a
conversion of the surface integral into one over the
volume of the body. However, as the physical fields in
the volume are not continuous with their respective
surface values, virtual fields have to be defined which
do. They need not be healthy fields, may violate the
Maxwell equations, or have a non-physical susceptibility.

We now derive Eqs~(\ref{emsf2}) by first considering a
solid body submerged in fluid and attached to a string.
Both the solid and fluid may be magnetizable
--- though differently, with the Maxwell stress $\Pi^{\rm
tot}_{ij}$ respectively denoted as $\Pi^{\rm s}_{ij}$ and
$\Pi^{\rm f}_{ij}$. One of the boundary conditions at the
interface is then
\begin{equation}\label{sBC}
\sigma^{\rm s}_{ij}+\Pi^{\rm s}_{ij}=\Pi^{\rm
f}_{ij}+\sigma^{\rm f}_{ij},
\end{equation}
where $\sigma^{\rm s}_{ij}\equiv$ $\partial u^{\rm tot}
/\partial u_{ij}$ is the elastic stress tensor of the
solid, given by deriving the thermodynamic energy with
respect to the elastic strain $u_{ij}$. In equilibrium,
$\nabla_j\sigma^{\rm s}_{ij}$ $=0$. Being a liquid,
$\sigma^{\rm f}_{ij}$ $\equiv0$ except where the surface
integral cuts across the string, which also has
elasticity we need to account for. Same as
Eqs~(\ref{54}), this boundary condition states the
continuity of the total momentum current $\sigma_{ij}+
\Pi_{ij}$. [The surface tension $\alpha$ is irrelevant
for a solid surface of given shape and therefore
eliminated.]

Each of the four terms of Eq~(\ref{sBC}) stands for a
surface force density. So the integral over the closed
surface of the solid body must yield the equation of
force equilibrium, between the gravitational,
electromagnetic and elastic force,
\begin{equation}\label{forceEquil}
{\cal F}^{\rm elast}_i+{\cal F}^{\rm grav}_i+{\cal
F}^{\rm elm}_i=0.
\end{equation}
The first term of Eq~(\ref{sBC}) vanishes, because
$\oint\sigma^{\rm s}_{ij}{\rm d}A_i$
$=\int\nabla_j\sigma^{\rm s}_{ij}{\rm d}^3r$ $=0$. The
last term yields the elastic force exerted by the string,
which is given by the normal component, ${\cal F}^{\rm
elast}_i=$ $-\oint\sigma^{\rm f}_{nn}{\rm d}A_i$, because
the string cannot sustain a shear stress, $\sigma^{\rm
f}_{tn}=0$. Clearly, this implies
\begin{equation}\label{kraft}
{\cal F}^{\rm grav}_i+{\cal F}_i^{\rm
elm}=\oint\triangle\Pi^{\rm tot}_{ij}\,{\rm d}A_j.
\end{equation}
Because $\oint\triangle\Pi^{\rm tot}_{ij}\,{\rm d}A_j=$
$\oint\triangle\Pi^{\rm tot}_{ij}n_j\,{\rm d}A=$ $\oint
\delta_{ik}\triangle\Pi^{\rm tot}_{kj}$ $n_j\,{\rm d}A=$
$\oint(t_it_k+n_in_k)\triangle\Pi^{\rm tot}_{kj}n_j\,{\rm
d}A=$ $\oint(n_i\triangle\Pi^{\rm tot}_{nn}+
t_i\triangle\Pi^{\rm tot}_{tn})\,{\rm d}A$, and because
$\triangle\Pi^{\rm tot}_{tn}$ vanishes identically, cf
Eq~(\ref{54x}), this reduces to
\begin{equation}\label{nKraft}
{\cal F}_i^{\rm elm}+{\cal F}_i^{\rm grav}=
\oint\triangle\Pi^{\rm tot}_{nn}\,{\rm d}A_i.
\end{equation}
Inserting Eq~(\ref{301}) for $\triangle\Pi^{\rm
tot}_{nn}$, the Gauss law is employed to find $\triangle
\oint K {\rm d}A_i=\triangle\int\nabla_i K{\rm d^3}r$
$=0$. As a result, only terms that depend explicitly on
fields are left. From these, the gravitational force is
found to be ${\pmb{\cal F}}^{\rm grav}=-\oint
gz\triangle\bar\rho\, {\rm d}{\bf A}$ $=-\int{\boldsymbol
\nabla}(\triangle\bar\rho gz){\rm d}^3r=$ $-g\hat{\bf
e}_z\triangle\int\!\rho\,{\rm d}^3r$, while the
electromagnetic force is found to be given by
Eq~(\ref{emsf2}).

Because Eq~(\ref{sBC}) is valid for any interface, and
not confined to the considered geometry, so are the
formulas Eqs~(\ref{emsf2}), (\ref{emsf}), (\ref{kraft}),
and (\ref{nKraft}). Clearly, there is an electromagnetic
surface force density of the given form whenever the
Maxwell stress is discontinuous, ie when the
magnetization (or polarization) changes abruptly.

To confirm this, consider a further example: Ferrofluid
contained in a nonmagnetic vessel, and the whole system
hung on a string in atmosphere. It is slightly more
complicated geometry, as there are two interfaces to
consider: ferrofluid-solid, and solid-air. Each has a
boundary condition of the type of Eq~(\ref{sBC}). We have
$\sigma^{\rm s}_{ij}= \sigma^{\rm a}_{ij}$ at the
solid-air interface, because the Maxwell stress is
continuous there, $\Pi^{\rm s}_{ij}=\Pi^{\rm a}_{ij}$. An
integration yields, as before, $\oint \sigma^{\rm s}_{nn}
\,{\rm d}A_i=$ $\oint \sigma^{\rm a}_{nn} \,{\rm d}A_i$
$=-{\cal F}_i^{\rm elast}$. The boundary at the
ferrofluid-solid interface is $\Pi^{\rm ff}_{ij}=
\Pi^{\rm s}_{ij}+\sigma^{\rm s}_{ij}$, Integrating over
the interface, we find $\oint(\Pi^{\rm ff}_{ij}- \Pi^{\rm
s}_{ij})\,{\rm d}A_i$ $+{\cal F}_i^{\rm elast}=0$, or
again that Eq~(\ref{kraft}), and therefore
Eqs~(\ref{nKraft}) and (\ref{emsf2}) to be valid.

In the literature~\cite{rz,LL8}, instead of
Eq~(\ref{emsf2}), the electromagnetic force is usually
given as
\begin{equation}\label{LLforce}
{\cal F}^{\rm elm}=\oint(H_iB_j-{\textstyle\int}B_k{\rm
d} H_k \delta_{ij}){\rm d}A_j.
\end{equation}
This expression may also be obtained by integrating over
Eq~(\ref{sBC}). First note $\oint(\sigma^{\rm s}_{ij}$
$+\Pi^{\rm s}_{ij}) {\rm d}A_j$
$=\int\nabla_j(\sigma^{\rm s}_{ij}$ $+\Pi^{\rm s}_{ij})
{\rm d}^3r$ $=0$, because the integrand, as the total
momentum flux density, has vanishing divergence in
equilibrium (neglect, for simplicity, gravitation).
Writing the force equilibrium as $\oint(\Pi^{\rm
f}_{ij}+\sigma^{\rm f}_{ij}) {\rm d}A_j$ $=0$, again
identifying ${\cal F}^{\rm elast}_i=$ $-\oint\sigma^{\rm
f}_{ij}{\rm d}A_j$, we find ${\cal F}^{\rm elm}_i=$
$-\oint \Pi^{\rm f}_{ij}{\rm d}A_j$ by comparison with
Eq~(\ref{forceEquil}). Now, inserting Eq~(\ref{2dB/dH})
for $\Pi^{\rm f}_{ij}$, and because $\oint K {\rm d}A_j$
$=0$, we obtain Eq~(\ref{LLforce}).

Although algebraically equivalent, Eqs~(\ref{emsf2}) and
(\ref{LLforce}) may behave quite differently when being
evaluated for a concrete geometry. For instance,
considering the force on a plate starting from
Eq~(\ref{LLforce}), it is not easy to find the result of
Eq~(\ref{emvf!}). This is because Eq~(\ref{LLforce}) is a
rather nonlocal expression, which does not even
explicitly depend on the susceptibility of the
magnetizable body.

The above lengthy discussion should not obscure the point
that these equilibrium surface forces are no more than
interpretation and visualization of boundary conditions,
especially Eqs~(\ref{301}, \ref{sBC}). In fact, when
considering the experiments in Chapter~\ref{exp},  we
shall usually employ the boundary conditions, to solidify
the algebraic, safe, albeit slightly more tedious
approach. But we shall also frequently point out how the
surface forces considered here would have yielded the
same results. The latter serves to demonstrate that these
forces represent a heuristic concept of considerable
power, and to give further reassurance, if one is needed,
of the validity of the formulas in this section.

\subsection{Thermodynamic Derivation
of the Stress}\label{ener}

It is worthwhile to rederive our central result, the
Maxwell tensor, Eq~(\ref{51}), via a more direct path,
though with a rather narrower focus and ensured range of
validity. This is done in the form of an expanded
Landau/Lifshitz consideration, separately for the
electric and magnetic contributions. We start from the
energy expression Eq~(\ref{rfe}), or from one of its
potentials, Eqs~(\ref{F}, \ref{tildeF}, \ref{G},
\ref{tildeG}), and shall take pains in eliminating the
flaws mentioned in the introduction.

Deforming an isolated system at given entropy, mass and
electric charge, the energy change is
\begin{equation}\label{51ab}
\delta U\equiv\delta\int u^{\rm tot}{\rm d}{V}=
-\oint(\triangle\Pi_{ik}^{\rm tot}\delta r_i){\rm d}A_k,
\end{equation} where the energy density $u^{\rm tot}$ is
given by Eq~(\ref{rfe}) with $\boldsymbol v$ $=0$.
Notations: $\delta r_i$ is the infinite\-simal (or
virtual) displacement of the surface, and ${\rm d}A_k
\equiv n_k{\rm d}A$ with ${\rm d}A$ the surface element
and $n_k$ the surface normal pointing outwards. The
validity of this equation is connected to
$-\triangle\Pi_{ik}^{\rm tot}$ being the electromagnetic
force density needed to deform the system, see
Eq~(\ref{kraft}). All gravitational terms are neglected
in this section.

If systems are considered for which the external stress
tensor vanishes, because neither material nor field lies
outside the considered volume, we may substitute
$\Pi_{ik}^{\rm tot}$  for $\triangle\Pi_{ik}^{\rm tot}$
in Eq~(\ref{51ab}). Only these systems will be considered
below. In the absence of fields, $\Pi_{ik}^{\rm tot}\to
P\delta_{ik}$ with $P$ uniform in thermodynamic
equilibrium, so $\delta U$ reduces to $-P\oint\delta
r_i{\rm d}A_i= -P\delta {V}$, or Eq~(\ref{31}). Note that
if the projection of the surface force $\Pi_{nn}^{\rm
tot}=\Pi_{ik}^{\rm tot}n_in_k$ is positive (ie positive
pressure if without field), the volume tends to expand,
if $\Pi_{nn}^{\rm tot}<0$, it tends to contract.

In simple geometries, if $u^{\rm tot}$ and $\Pi_{ik}^{\rm
tot}\delta r_i$ are uniform, Eq~(\ref{51ab}) reduces to
\begin{equation}\label{51aa}
\delta U= \delta (u^{\rm tot}{V}) = u^{\rm tot}\delta
{V}+{V}\,\delta u^{\rm tot}= -A_k\Pi_{ik}^{\rm tot}\delta
r_i.
\end{equation} Since $u^{\rm tot}$ is known,
we shall evaluate $A_k\Pi_{ik}^{\rm tot}\delta r_i$ while
taking $A_k$ and $\delta r_i$ to point in all three
directions, perpendicular and parallel to the field, and
hereby obtain all nine components of $\Pi_{ik}$.

\subsubsection{The Electric Part}
We consider a parallel-plate capacitor that is filled
with a dielectric medium. Denoting the three linear
dimensions of the capacitor as $x, y, z$, with $x\ll
y,z$, the six surfaces $\,\,S_x^\pm, \,\,S^\pm_y,
\,\,S_z^\pm$ (with the outward pointing normal
$\,\,\pm\hat{\bf e}_x,
 \,\,\pm\hat{\bf e}_y,$ $\,\,\pm\hat{\bf e}_z$) have the areas
$A_x=yz,\ A_y=xz,\ A_z=xy$, respectively; and the volume
is ${V}=xyz$. Taking the two metal plates as $S_x^\pm$,
the electric fields $E, D$ are along $\hat{\bf e}_x$, see
Fig 1. (We neglect the small stray fields at the edges.)
The capacitor is placed in a va\-cuum, so there is
neither electric field nor material outside.

\begin{figure}
\begin{center}
\includegraphics[width=0.8\columnwidth]{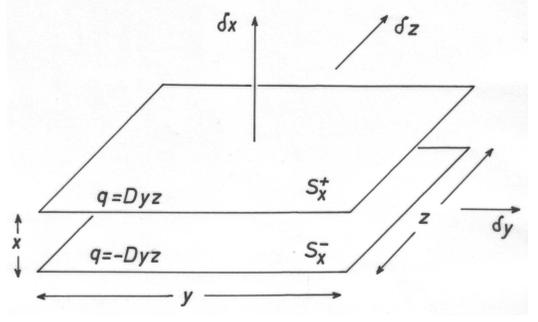}
\end{center}
\caption{Metal plates at $S_x^+$ and $S_x^-$ with
constant charge, dielectric medium sandwiched between
them. Displacing $S_x^+$ by $\delta x$ or $\delta z$
respectively decompresses and shears the
medium.}\label{Fig1}
\end{figure}

We now successively displace the three surfaces $S^+_x$,
$S^+_y$, $S_z^+$, in all three directions, $\delta
r_i=\delta x, \delta y, \delta z$ (which is why the
capacitor has to be finite), while holding constant the
quantities: entropy $s{V}$, masses $\rho_\alpha{V}$, and
electric charges $q=\pm DA_x$ (from
${\boldsymbol\nabla}\cdot{\bf D}=\rho_{\rm e}$). Because
of the simple geometry, Eq~(\ref{51aa}) holds and will be
evaluated.

First, take the surface to be displaced as $S_x^+$. When
the displacement is $\delta x$, we have
\begin{equation}\label{101}
 \delta{V}=A_x\delta x,\ \delta s/s =
 \delta \rho_\alpha/\rho_\alpha=-\delta
x/x,\ \delta D=0; \end{equation}   and we have
$\delta{V}$, $\delta s$, $\delta\rho_\alpha$, $\delta
D=0$ if the displacement is $\delta y$ or $\delta z$
(implying a shear motion of $S_x^+$). Inserting all three
into Eq~(\ref{51aa}), we obtain
\begin{eqnarray}\label{55}
 \Pi_{xx}^{\rm tot}\delta x=(Ts+ \xi_\alpha\rho_\alpha-
 u^{\rm tot})\delta
x,\quad \Pi_{yx}^{\rm tot}=\Pi_{zx}^{\rm
tot}=0.\quad\quad
\end{eqnarray}
If the surface is $S_z^+$ and the displacement $\delta
z$, we have $\delta{V}=A_z\delta z$ and $\delta s/s$,
$\delta \rho_\alpha/\rho_\alpha$, $\delta D/D=-\delta
z/z$. If the displacement is $\delta x$ or $\delta y$, we
have $\delta{V}$, $\delta s$, $\delta \rho_\alpha$,
$\delta D=0$. Hence
\begin{equation}
\Pi_{zz}^{\rm tot}\delta
z=(Ts+\xi_\alpha\rho_\alpha+E_xD_x-u^{\rm tot})\delta z,
\label{55a}\end{equation} and $\Pi_{xz}^{\rm
tot}=\Pi_{yz}^{\rm tot}=0$. As the directions $\hat{\bf
e}_y$ and $\hat{\bf e}_z$ are equivalent, we know without
repeating the calculation that a displacement of $S_y^+$
yields $\Pi_{zz}^{\rm tot}=\Pi_{yy}^{\rm tot}$ and
$\Pi_{xy}^{\rm tot}=\Pi_{zy}^{\rm tot}=0$. (The term
$E_xD_x$ is a result of the metal plates being squeezed,
compressing the surface charges, $\delta q/q=\delta
D/D=-\delta z/z$. The compressibility of the metal is
taken to be infinite. Otherwise, it would contribute an
additional term in the stress tensor that we are not
interested in here.)

These considerations have yielded all nine components of
$\Pi_{ik}^{\rm tot}$ for a special coordinate system.
Because the stress tensor of Eq~(\ref{51}), for ${\bf
D,E\|}\hat{\bf e}_x$ and ${\bf B}$, $\boldsymbol v$ $=0$
produces exactly these components, it is the correct,
coordinate-independent expression.

This conclusion may appear glib, but is in fact quite
solid. Consider first a vector: If two vectors are shown
to be equal in a special coordinate system, we know that
they remain equal in any other system -- as long as we
are sure that they are indeed vectors. The same also
holds for tensors. (The careful reader may notice an
ambiguity with respect to the off-diagonal part, as both
$E_iD_k$ and $E_kD_i$ yields the same nine components
derived here. Fortunately, there is no difference between
these two expressions, because $\bf E\times D=0$ for
${\bf B},\mbox{\boldmath$v$}=0$, cf Eq (\ref{64}).) This
concludes a thermodynamic derivation of the electric part
of the Maxwell stress tensor.

Substituting the dielectric medium with vacuum,
$\Pi_{xx}^{\rm tot}$, $\Pi_{zz}^{\rm tot}$ reduce to $\mp
e^2/2$, respectively, implying a tendency to contract
along $\hat{\bf e}_x$ and expand along $\hat{\bf e}_y$
and $\hat{\bf e}_z$ --- as it should in the considered
case: The differently charged plates want to come closer,
while the charge in each plate would like to expand. If
the medium is one with negligible susceptibilities,
$\Pi_{xx}^{\rm tot},\Pi_{zz}^{\rm tot}$ are given as
$P\mp e^2/2$, so the same electric force must now contend
with the pressure in deforming the system.

Consider the same capacitor, now held at a constant
voltage $\phi$. The modified system must lead to the same
stress tensor, because the stress tensor is a local
expression and may depend only on the local field. The
calculation is rather similar, though one needs to
replace $u^{\rm tot}$ in Eq~(\ref{51aa}) with the
potential $\tilde{u}\equiv u^{\rm tot}-{\bf E}\cdot{\bf
D}$, as the system is no longer electrically isolated.
And the constraint for $E$ as the new variable is
$Ex=\phi$ (which replaces $DA_x=q$). Connecting the
capacitor in addition to a heat bath necessitates the
employment of the free energy, $\tilde F=u^{\rm
tot}-Ts-{\bf E}\cdot{\bf D}$, and changes the constraint
(from constant $s{V}$) to $\delta T=0$. ($\tilde F$ is
the potential used in~\cite{LL8}.) For the explicit
calculation please cf the magnetic case below,
Eqs~(\ref{Ftilde}, \ref{Ftilde2}, \ref{101mag},
\ref{55mag}, \ref{55amag}), with the replacement $\bf
B\to D,\, H\to E$ implemented. The final result is the
same, given by Eq~(\ref{51}).

\subsubsection{The Magnetic Part}\label{mag}
To obtain the magnetic part of the stress tensor,
consider a rod along $\hat{\bf e}_x$, of square cross
section, made of a magnetizable material and placed in a
vacuum. The surfaces $S^\pm_y,S_z^\pm$ are covered with a
sheet of metal that carries a current $J\perp\hat{\bf
e}_x$. With $A_x\ll A_y,A_z$, the magnetic field will be
essentially along $\hat{\bf e}_x$ and confined to the
interior of the rod, see Fig 2. So again, there is
neither field nor material outside.

\begin{figure}
\begin{center}
\includegraphics[width=0.4\columnwidth]{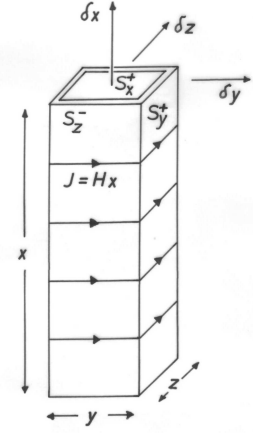}
\end{center}
\caption{A magnetic system with constant current
$J\perp\hat{\bf x}$, enforced by an external battery, not
shown. Again, it is deformed by displacing say $S_x^+$
along $\hat{\bf x}$ or $\hat{\bf z}$. }\label{Fig2}
\end{figure}

If the system is isolated, the metal needs to be
superconducting to sustain the current, and the
constraint on the variable $B$ during a deformation is
constant flux, $BA_x=\Phi$. (Compare this to the isolated
electric case with $DA_x=q$.) If the current $J$ is held
constant by a battery, the constraint is $Hx=J/c$ (from
${\boldsymbol\nabla}\times{\bf H}={\bf j_e}/c$ and
analogous to $Ex=\phi$), and the attendant potential is
$\tilde{u}\equiv u^{\rm tot}-{\bf H}\cdot{\bf B}$.

The calculation of the isolated magnetic case
repeats the isolated electric one, with all
equations -- both the displayed
Eqs~(\ref{51aa},\ref{101},\ref{55},\ref{55a}) and
the others around these -- remaining valid after the
displacement $\bf D\to B,\, E\to H$ has taken place.
(When considering compressional displacements, the
elastic energy of the metal sheet is again
neglected.)

We shall now consider the rod being deformed at
constant current and temperature, so the energy
needed to deform the system is
\begin{eqnarray}\label{Ftilde}
\delta (\tilde F{V})=\tilde F\delta {V}+{V}\,\delta
\tilde F= -A_k\Pi_{ik}^{\rm tot}\delta r_i,\\ \delta
\tilde F=-s\delta T+\xi_\alpha\delta
 \rho_\alpha -{\bf B}\!\cdot\delta {\bf H},
 \label{Ftilde2}\end{eqnarray}
where $\tilde F=u^{\rm tot} -Ts-{\bf H\!\cdot B}$. We
again successively displace the three surfaces $S^+_x,
S^+_y, S_z^+$, in all three directions, $\delta
r_i=\delta x, \delta y, \delta z$, while holding constant
the quantities: temperature, masses, and the current, ie
under the conditions, $\delta T=0$, $\delta
(\rho_\alpha{V})=0$ and $\delta(Hx)=0$.

The first surface to be displaced is $S_x^+$. When the
displacement is $\delta x$, we have
\begin{equation}\label{101mag}
 \delta{V}=A_x\delta x,\ \delta H/H =
 \delta \rho_\alpha/\rho_\alpha=-\delta x/x;
\end{equation}   and we have
$\delta{V}$, $\delta\rho_\alpha$, $\delta H=0$ if the
displacement is $\delta y$ or $\delta z$ (implying a
shear motion of $S_x^+$). Inserting these into
Eq~(\ref{Ftilde}), we obtain
\begin{eqnarray}\label{55mag}
\Pi_{xx}^{\rm tot}\delta x=(Ts+ \xi_\alpha\rho_\alpha-
u^{\rm tot})\delta x,\ \Pi_{yx}^{\rm tot}=\Pi_{zx}^{\rm
tot}=0.\qquad
\end{eqnarray}
If the surface is $S_z^+$ and the displacement $\delta
z$, we have $\delta{V}=A_z\delta z$, $\delta H=0$, and
$\delta \rho_\alpha/\rho_\alpha=-\delta x/x$. If the
displacement is $\delta x$ or $\delta y$, we have
$\delta{V}$, $\delta \rho_\alpha$, $\delta H=0$. Hence
\begin{equation}
\Pi_{zz}^{\rm tot}\delta
z=(Ts+\xi_\alpha\rho_\alpha+H_xB_x-u^{\rm tot})\delta z,
\label{55amag}
\end{equation} and $\Pi_{xz}^{\rm
tot}=\Pi_{yz}^{\rm tot}=0$. Since the directions
$\hat{\bf e}_y$ and $\hat{\bf e}_z$ are equivalent, we
have $\Pi_{zz}^{\rm tot}=\Pi_{yy}^{\rm tot}$ and
$\Pi_{xy}^{\rm tot}=\Pi_{zy}^{\rm tot}=0$.

This consideration has yielded all nine components of
$\Pi_{ik}^{\rm tot}$ for a special coordinate system.
Because the stress tensor of Eq~(\ref{51}), for ${\bf
B,H}\|\hat{\bf e}_x$ and ${\bf D}=0$ produces these
components, it is the correct, coordinate-independent
expression. This concludes the thermodynamic derivation
of the magnetic part of the Maxwell stress tensor.

Although we have only employed the potentials $F$ and
$\tilde F$ above, it should be clear by now that we could
just as well have employed $G$ and $\tilde G$ from
Eqs~(\ref{G},\ref{tildeG}), assuming that the polarizable
medium is connected to particle reservoirs, which keep
the chemical potentials $\xi_\alpha$ constant. This
changes the constraint from $\delta\rho_\alpha/
\rho_\alpha =-\delta x/x,\,-\delta y/y$ to simply
$\delta\xi_\alpha=0$. The derived expression for the
stress tensor remains unchanged.

As a stand-alone, the thermodynamic consideration of the
last two sections gives us a fairly clear idea on the
form of the macroscopic Maxwell stress tensor
$\Pi_{ik}^{\rm tot}$, in equilibrium, for $v\equiv0$, and
with either the electric or the magnetic field present.
The most important information it withholds is that about
$g_i^{\rm tot}$, without which $\Pi_{ik}^{\rm tot}$ is
not unambiguously defined at finite frequencies. Also,
$g_i^{\rm tot}$ is needed to complete and close the set
of differential equations given previously, which alone
is capable of providing a consistent and comprehensive
description of polarizable systems. In addition, relying
solely on this consideration, one would perhaps need to
be more careful with the stray fields, especially when
thinking about the possibilities of terms such as
$\nabla_iE_j, \nabla_iH_j$.

Rosensweig has also considered and derived the magnetic
part of the stress tensor in~\cite{rz}. His result is the
same as the one here, though his algebra is rather more
complicated. More problematically, one of his basic,
starting assumptions does not hold up: His geometry is a
slab with $A_y, A_x\ll A_z$ and current-carrying wires
along the surfaces $S^\pm_y,S_z^\pm$, see his Fig 4.1.
The winding of the wires is oblique, the currents flow
along $\pm\hat{\bf e}_y$ in the two larger plates
$S_z^\pm$, but has a component along $\pm\hat{\bf e}_x$
in the two narrow side walls $S^\pm_y$ -- take them to be
along $\pm\hat{\bf m}$, a vector in the $xz$-plane.
Rosensweig maintains that the resultant field is uniform
and perpendicular to the surface given by the winding, ie
by $\hat{\bf e}_y$ and $\hat{\bf m}$.

Unfortunately, the field is neither uniform nor mainly
oblique, rendering large parts of the ensuing
consideration invalid. First the qualitative idea: If the
two much larger plates $S_z^\pm$ were infinite, the field
would be strictly parallel to $\hat{\bf e}_x$. This basic
configuration should not change much if the plates are
made finite, and supplemented with the two narrow side
walls  $S^\pm_y$ -- irrespective of the currents'
direction there. This argument is born out by a
calculation to superpose the fields from various portions
of the currents. First, divide all currents along
$\pm\hat{\bf m}$ into two components, along  $\pm\hat{\bf
e}_z$ and $\pm\hat{\bf e}_x$. Next, combine the first
with the currents along $\pm\hat{\bf e}_y$, such that the
four sections of the four surfaces form a closed loop at
the same x-coordinate. The resultant field of all loops
is clearly the main one, and strictly along $\hat{\bf
e}_x$. The leftover currents are those at $S^\pm_y$ along
$\pm\hat{\bf e}_x$ and their total effect is a small
dipole field.

\subsection{Ambiguous Notations}\label{MN}
In this section, we shall consider some frequently
employed expressions and notations that we shall see are
rather misleading, and therefore perhaps best avoided. On
the other hand, many of us have so accustomed ourselves
to this notation that we tend to think along the
categories it provides. But even then, or especially
then, one should welcome the opportunity to realize all
its pitfalls which, as discussed in the introduction, lie
mainly in the ambiguity of the zero-field pressure, and
of $\int{\bf M}\cdot{\rm d}{\bf H}$. The expressions one
arrives at are for instance $(P+ {\textstyle
\frac{1}{2}}H^2)\delta_{ik} -H_iB_k$ for the stress, or
$M_i\nabla H_i$ for the Kelvin force. Scrutinizing their
derivation, we shall in addition conclude that these
formulas are only valid for small susceptibility
$\chi^{\rm m}\ll1$, or $M\ll H$ --- a range of validity
dramatically smaller than taken for granted usually. So
if $\chi^{\rm m}$ is of order unity, and $M\approx H$,
 as is frequently the case in strongly polarizable
systems such as ferrofluids, these formulas are invalid.

\subsubsection{Different Zero Field Pressures}
\label{TvsS} We shall take an approach here that is
somewhat broader than usual, and simultaneously work with
different potentials. This will lead to rather flagrant
appearing contradictions, the understanding of which
should lend us a sharpened view of the adopted notation.
To simplify the formulas, we shall neglect gravitation,
and consider the incompressible limit of ferrofluids,
$\rho_\alpha, \xi_\alpha$ $\to\rho_1, \xi^{\rm in}_1$, cf
section~~\ref{incomp}. (Replacing $\rho_1, \xi^{\rm
in}_1$ with $\rho, \xi$ yields the results for a
one-component, compressible liquid.)

We start by separating the energy $u^{\rm
tot}(s,\rho_1,{\bf B})$, the free energy $F(T,\rho_1,{\bf
B})$, and the potentials $\tilde F(T,\rho_1,{\bf H})$,
$G(T,\xi^{\rm in}_1,{\bf B})$, and $\tilde G(T,\xi^{\rm
in}_1,{\bf H})$, cf Eqs~(\ref{F} - \ref{tildeG}), into
their zero-field and electromagnetic contributions,
\begin{equation}\label{83}
u^{\rm tot}=u(0)+u_{\rm em}({\bf B}),\quad F=F(0)+F_{\rm
em}({\bf B}),
\end{equation}
with analogously defined $G_{\rm em}({\bf B})$, $\tilde
F_{\rm em}({\bf H})$, $\tilde G_{\rm em}({\bf H})$.
Adhering to convention, we write the potentials that are
functions of $\bf B$ as
\begin{equation}\label{83a}
u_{\rm em},\,F_{\rm em},\,G_{\rm em}={\textstyle\int}{\bf
H}\cdot{\rm d}{\bf B}=
{\textstyle\frac{1}{2}}B^2-{\textstyle\int}{\bf
M}\cdot{\rm d}{\bf B},
\end{equation}
and the potentials with tilde that are functions of $\bf
H$ as
\begin{equation}\label{833}
\tilde G_{\rm em},\tilde F_{\rm em}=
-{\textstyle\int}{\bf B}\cdot{\rm d}{\bf H}=
-{\textstyle\frac{1}{2}}H^2-{\textstyle\int}{\bf
M}\cdot{\rm d}{\bf H}.
\end{equation}
Note that the respective integral is to be taken at
constant $s,\rho_1$ for $u_{\rm em}$, at constant
$T,\rho_1$ for $F_{\rm em}$, $\tilde F_{\rm em}$, and at
constant $T,\xi^{\rm in}_1$ for $G_{\rm em}$, $\tilde
G_{\rm em}$. So $u_{\rm em},\,F_{\rm em},\,G_{\rm em}$
are in fact not equal, neither are $\tilde G_{\rm em}$,
$\tilde F_{\rm em}$. The situation is similar to
evaluating $\int T{\rm d}s$ when considering a Carnot
process: Since $T=T(s,\rho)$, we need to specify what the
density $\rho$ does when $s$ varies. [We have already
introduced $-\tilde G_{\rm em}=$
${\textstyle\frac{1}{2}H^2} +{\textstyle\int_{\rm
Eq}}M_i{\rm d}H_i$ and $-\tilde F_{\rm em}=$ ${\textstyle
\frac{1}{2}H^2} +{\textstyle\int_{\rm uni}}M_i{\rm d}H_i$
in Eqs~(\ref{mbe1}, \ref{sn3}), given each a specifying
index, and discussed why the difference between these
expressions is irrelevant as long as {\sc lcr} holds, cf
Eq~(\ref{ti2}).]

To evaluate the stress $\Pi_{ik}$,  Eq~(\ref{51}), in
terms of these potentials, we start with its diagonal
part $-\tilde G$, given respectively as
\begin{eqnarray}
\label{833f}-\tilde G=
(s\textstyle{\frac{\partial}{\partial s} + \rho_1
\frac{\partial} {\partial\rho_1}} - 1)\, u^{\rm
tot}=P(s,\rho_1)+ {\textstyle\frac{1}{2}}H^2\quad
\nonumber\\ -{\textstyle\frac{1}{2}}M^2
+{\textstyle\int{\rm d}{\bf
B}\!\cdot\!(1-s\frac{\partial}{\partial
s}-\rho_1\frac{\partial}{\partial\rho_1})} {\bf M},\quad
\\ \label{833g}-\tilde G={\textstyle( \rho_1\frac{\partial}
{\partial\rho_1}-1)}\, F=P(T,\rho_1)+
{\textstyle\frac{1}{2}}H^2\quad\nonumber\\
-{\textstyle\frac{1}{2}}M^2+ {\textstyle\int{\rm d}{\bf
B} \cdot(1-\rho_1\frac{\partial} {\partial\rho_1})}{\bf
M},\quad
\\ \label{833h}-\tilde G={\textstyle(\rho_1\frac{\partial}
{\partial\rho_1}-1)}\, \tilde
F=P(T,\rho_1)+{\textstyle\frac{1}{2}}H^2\quad \nonumber\\
+{\textstyle\int{\rm d}{\bf
H}\cdot(1-\rho_1\frac{\partial} {\partial\rho_1})}{\bf
M},\quad\\ \label{833i}-\tilde G=P(T,\xi^{\rm in}_1)+
{\textstyle\frac{1}{2}}H^2 + {\textstyle\int}{\bf M
\cdot{\rm d}H},\quad
\end{eqnarray}
where depending on the variables, the zero-field
pressures are 
$P(s,\rho_1)=$ $(s\frac{\partial}{\partial s}+\rho_1
\frac{\partial} {\partial\rho_1} - 1)u(0)$,
$P(T,\rho_1)=$ $(\rho_1\frac{\partial}
{\partial\rho_1}-1)F(0)$, and $P(T,\xi^{\rm in}_1)=$
$-\tilde G(0)$, cf Eq~(\ref{31}) with $u(0)=u^{\rm M}$,
$v_i=0$. These four differing $\tilde G$, and with them
the attendant stress $\Pi_{ik}^{\rm tot}=-\tilde
G\delta_{ik}-H_iB_k$ certainly look disturbingly
contradictory. Yet all must be equal. Equating
Eq(\ref{833h}) with (\ref{833i}), we find $P(T,\xi^{\rm
in}_1)=$ $P(T,\rho_1)$ $-\int\rho_1\partial{\bf
M}/\partial\rho_1\!\cdot{\rm d}{\bf H}$, forcing the
conclusions that the zero field pressure $P$ is a
quantity that depends on the variables chosen, and that
their difference is of order magnetic field squared. This
means especially that only one of the pressures may be
field-independent under given circumstances.

The pressure all textbooks (including~\cite{rz,LL8}) take
to be field independent is $P(T,\rho)$, equivalent to
$P(T,\rho_1)$ in our case. This is the pressure that one
would measure in the absence of field, for given
temperature and density. The gradient of the remaining,
field dependent terms, $\nabla_j[\Pi^{\rm tot}_{ij}$
$-P(T,\rho_1)\delta_{ij}]$, cf Eq~(\ref{833h}), are then
interpreted as the electromagnetic force, referred to as
ponderomotive or Helmholz force. But all this is
obviously only valid if the temperature and density are
indeed kept constant when the field is switched on. Under
adiabatic circumstances, when $s$ is kept constant
instead of $T$, the pressure $P(s,\rho_1)$ is the one
that is field independent. Consequently, the above
Helmholz force is no longer the correct expression for
the electromagnetic force. Rather, one must take the
field dependent terms from Eq~(\ref{833f}) instead.

The ambiguity of $P$ arises from the fact that the
dependent variables change when the field is switched on.
For instance, the chemical potential, $\xi^{\rm in}_1=$
$\partial\tilde F(0)/\partial \rho_1+$ $\partial\tilde
F_{\rm em}/\partial \rho_1=$ $\xi^{\rm in}_1(0) +\Delta
\xi^{\rm in}_1$, changes by the amount $\Delta \xi^{\rm
in}_1=$ $\partial\tilde F_{\rm em}/\partial \rho=$
$-\int{\rm d}{\bf H}\cdot\frac{\partial}
{\partial\rho_1}{\bf M}$. Therefore
\begin{eqnarray}
P(T,\xi^{\rm in}_1)=P(T,\xi^{\rm
in}_1\mbox{\footnotesize(0)})+{\textstyle \frac{\partial
P}{\partial \xi^{\rm in}_1}}\Delta\xi^{\rm in}_1\nonumber
\\= P(T,\rho_1)-{\textstyle\int {\rm d}{\bf
H}\cdot\rho_1\frac{\partial} {\partial\rho_1}}{\bf
M},\label{834xx}
\end{eqnarray}
explaining the difference between Eqs~(\ref{833h}) and
(\ref{833i}). The ill behavior of $P$ takes one by
surprise, as proper thermodynamic expressions do not
depend on the variables chosen. This stems from the fact
that $P$ is not a thermodynamically defined quantity in
the presence of fields: It does not appear in
Eq~(\ref{42}), as all bona fide thermodynamic variables
do. Since Eq~(\ref{51x}) or (\ref{51}) consist only of
quantities which do appear in Eq~(\ref{42}) -- both with
or without field -- the stress tensor $\Pi_{ik}^{\rm
tot}$ given there holds for any set of variables. These
are good reasons to make this the expression of choice.

Frequently, a further approximation is employed for the
diagonal part of the stress. In dilute, one-component
systems, the magnetization is usually proportional to the
density, or $M=\rho(\partial M/\partial\rho)$. Similarly,
in magnetically dilute ferrofluids, we may assume that
the magnetization is proportional to the particle density
$\rho_1$, $M= \rho_1(\partial M/\partial\rho_1)$.
Inserting this into Eqs~(\ref{833g}) and (\ref{833h}), we
find, respectively,
\begin{equation}\label{834x}
-\tilde G=P+ {\textstyle\frac{1}{2}}(H^2-M^2),
\quad-\tilde G=P+ {\textstyle\frac{1}{2}}H^2.
\end{equation}
The difference between these two expressions are not due
to a different set of variables, as $P$ is a function of
$T$ and $\rho_1$ in both cases. So we are dealing with a
different pitfall here, one that we shall discuss in
details in section~\ref{pond} below. The gist of it is:
When we assume $M\sim\rho_1$, this is meant as an
approximation, implying the neglect of square and higher
order terms $\sim\rho_1^2,\rho_1^3\cdots$. But
consistency then dictates that we must neglect all higher
order terms, including $M^2\sim\rho_1^2$. This implies
especially that the dilute limit is only given if $M\ll
H$ and $\chi^{\rm m}\ll1$ hold. Therefore, the term
$\frac{1}{2}M^2$ in Eqs~(\ref{834x}) must be neglected,
and the popular form for the Maxwell stress tensor,
$\Pi_{ik}=(P+ \frac{1}{2}H^2) \delta_{ik}-H_iB_k$, is to
be taken with a large grain of salt, as it is valid only
for $M\ll H$, and quite useless if $M\approx H$, or
$\chi^{\rm m}\approx1$.

\subsubsection{Magnetic Bernoulli Equation}
\label{bernoulli}

The magnetic Bernoulli equation by Rosensweig is a very
useful relation. It has been extensively employed in his
book~\cite{rz}, and in the literature on ferrofluids. We
shall include the variation of concentration, which he
did not consider, and in addition, free this relation
from the ambiguous notation of the last section
\ref{TvsS}, in which it is given. The point is, the
information in the magnetic Bernoulli equation is
contained in Eq~(\ref{dB/dH}). Combine it with
Eq~(\ref{833i}) to yield
\begin{equation}\label{mbe}
{\boldsymbol\Delta}[P(T,\xi^{\rm in}_1)+\bar\rho g z]=0,
\end{equation}
an expressions of the magnetic Bernoulli equation if the
system is in equilibrium with respect to particle
distribution. (Note that of all the
${\textstyle\int}M_i{\rm d}H_i$ in
Eqs~(\ref{833f}-\ref{833i}), only that in Eq~(\ref{833i})
has the same variables as ${\textstyle\int_{\rm
Eq}}M_i{\rm d}H_i$, hence these two expressions cancel
each other.) Before arriving at total equilibrium, and as
long as the concentration is uniform, we may start from
Eq~(\ref{sn2}) and combine it with Eq~(\ref{833h}). The
result is
\begin{equation}\label{mbe2}
  \mbox{\boldmath$\Delta$}
[P(T,\rho_1)+\rho g z-\rho_1
{\textstyle\frac{\partial}{\partial\rho_1}\int} M_i{\rm
d}H_i]=0.
\end{equation}
Substitute $\rho$ for $\rho_1$ to arrive at the
expression as given by Rosen\-sweig.

The velocity dependent terms in the original  magnetic
Bernoulli equation have not been included here, because
considerations of mass currents in ferrofluids need to
include viscosities. Besides, some of the velocity
dependent terms in the stress tensor, Eq~(\ref{51x}), are
missing in~\cite{rz}.

\subsubsection{Kelvin and Helmholtz Force}\label{pond}

As in section \ref{TvsS} above, we may separate out the
zero-field pressure $P$ from the hydrodynamic bulk force
density of Eq~(\ref{f}), $f^{\rm bulk}=$ $s\nabla
T+\rho_\alpha\nabla\xi_\alpha$. (We again neglect
gravitation, and consider a stationary medium, $v=0$.)
The remaining terms are referred to as the ponderomotive
force ${\bf f}^{\rm P}$,
\begin{eqnarray}
\label{86a}f^{\rm bulk} \equiv{\boldsymbol\nabla}
P(T,\rho_\alpha) -{\bf f}^{\rm P},\\
 \label{86}{\bf f}^{\rm P}= (\partial F_{\rm em}/
 \partial T){\boldsymbol\nabla} T-\rho_\alpha{\boldsymbol
  \nabla}(\partial F_{\rm
em}/\partial\rho_\alpha)
\\ ={\boldsymbol\nabla}(F_{\rm em}-\rho_\alpha \partial F_{\rm em}
/\partial\rho_\alpha)-H_i{\boldsymbol\nabla} B_i.
\label{nl1}
\end{eqnarray}
[The relation ${\boldsymbol\nabla} F_{\rm em}= (\partial
F_{\rm em}/
\partial T)\cdot{\boldsymbol\nabla} T +(\partial F_{\rm em}/\partial
\rho_\alpha)\cdot{\boldsymbol\nabla} \rho_\alpha
+H_i{\boldsymbol\nabla} B_i$ was used for the last equal
sign.]

In this section, we shall only employ the two potentials
$\tilde F(T,\rho_\alpha, H_i)$ and $F(T,\rho_\alpha,
B_i)$. So the zero field pressure depends on temperature
and densities, and the derived ponderomotive force is
valid only under isothermal condition of constant
densities, cf the discussion of section~\ref{TvsS}.

We now proceed to derive the Kelvin force by
incorporating some specific simplifications. For {\sc
lcr}, $-\partial F_{\rm em}/\partial T= \frac{1}{2}
H^2\partial\chi^{\rm m}/\partial T$, and similarly
$\partial F_{\rm em}/\partial\rho_\alpha$, so ${\bf
f}^{\rm P}$ can be cast as
\begin{equation}\label{87}
 {\bf f}^{\rm P}={\boldsymbol\nabla}
 (\textstyle\frac{1}{2}H^2
\rho_\alpha\partial\chi^{\rm m}/\partial\rho_\alpha)
-\textstyle\frac{1}{2}H^2{\boldsymbol\nabla}\chi^{\rm m},
\end{equation}
usually referred to as the Helmholtz force in the
literature. If the system is magnetically dilute, we have
$\chi^{\rm m}\sim\rho$ in a one-component gas, and
$\chi^{\rm m}\sim\rho_1$ for ${\rm d}\rho=$ $\gamma{\rm
d}\rho_1$ in ferrofluids (with no additional dependence
on $\rho$). Both imply $\rho_\alpha\partial\chi/\partial
\rho_\alpha= \chi$. Then the ponderomotive force reduces
to $\frac{1}{2}\chi^m{\boldsymbol\nabla} (H^2)$, or
\begin{equation} \label{888}
{\bf f}^{\rm P}=M_i{\boldsymbol\nabla} H_i,\quad {\bf
M}=\chi^{\rm m}{\bf H}.
\end{equation}
Assuming in addition a static field:
$\bf{\boldsymbol\nabla} \times H=0$, the force ${\bf
f}^{\rm P}$ may also be written as $(M_i\nabla_i){\bf
H}$, a form one often encounters.

This seems a satisfactory derivation of the Kelvin force,
as both the linear constitutive relation and the
proportionality to the density are frequently valid
approximations. Unfortunately, the obviously different
expression
\begin{equation}\label{88a}
{\bf f}^{\rm P}= M_i{\boldsymbol\nabla} B_i
\end{equation}
may be obtained by essentially the same derivation. We
start by defining a slightly different susceptibility:
${\bf M}=\tilde\chi^{\rm m}{\bf B}$. Although different
from the more usual convention of Eq~(\ref{888}), the new
susceptibility is undoubtedly physically equivalent to
the old one, and we have no a priori reason to prefer
either. Both susceptibilities are related via
$\tilde\chi^{\rm m}=1-1/\mu =\chi^{\rm m}/(1+\chi^{\rm
m})$, or ${\rm d}\tilde\chi^{\rm m}={\rm d}\chi^{\rm
m}/(\mu)^2$. So Eq~(\ref{87}) may be rewritten as
\begin{equation}\label{87a}
{\bf f}^{\rm P}={\boldsymbol\nabla}
(\textstyle\frac{1}{2} B^2 \rho_\alpha
\partial\tilde\chi^{\rm m}/\partial\rho_\alpha)
-\textstyle\frac{1}{2}B^2{\boldsymbol\nabla}\tilde\chi^{\rm
m}.
\end{equation}
This time, assuming $\tilde\chi^{\rm m}$ is proportional
to one of the densities, we obtain Eq~(\ref{88a}).

Which is the correct one, Eq~(\ref{888}) or (\ref{88a})?
Since Eq~(\ref{87})and (\ref{87a}) are algebraically
equivalent, the difference must stem from the two
assumptions, $\chi^{\rm m}\sim\rho$ and $\tilde\chi^{\rm
m}= \chi^{\rm m}/(1+\chi^{\rm m})\sim\rho$. Reviewing the
above derivations, it is obvious that if one of the two
assumptions were {\em strictly} correct, the other would
be wrong, and only the associated force expression is
applicable. Generically, however, both $\chi^{\rm m}$ or
$\tilde\chi^{\rm m}$ are power series of $\rho$. Assuming
either susceptibility is linear in $\rho$ is only an
approximation. And consistency dictates that all
quadratic terms are then to be discarded. This implies
(i) in the dilute approximation, both susceptibilities
may only retain the term linear in $\rho$ and are
therefore equal; (ii) all other terms quadratic in the
two susceptibilities are also to be discarded, as
$(\chi^{\rm m})^2,(\tilde\chi^{\rm m})^2 \sim\rho^2\to0$.
We conclude: the Kelvin force is only valid for
$\chi^{\rm m}\ll1$ or $M\ll H$, to linear order in
$\chi^{\rm m}$ or $M$. But then $M_i{\boldsymbol\nabla}
B_i$ and $M_i{\boldsymbol\nabla} H_i $ are equally valid.
The same of course also holds for $P_i{\boldsymbol\nabla}
D_i$ and $P_i{\boldsymbol\nabla} E_i $ in the electric
case.

The following example aims to illustrate this
conclusion~\cite{kelvin}. Consider a thin slab of
ferrofluid exposed to a homogeneous, external magnetic
field $B$, oriented normal to the slab. An enforced
temperature gradient within the slab ensures an
inhomogeneous susceptibility, $\chi^{\rm m}(T)$. The
internal $B$-field is uniform, but not the internal
$H$-field, as $H=B/(1+\chi^{\rm m})$. The ponderomotive
force of Eq~(\ref{88a}) is zero in the given
circumstance, but not that of Eq~(\ref{888}), which
yields $M_i{\boldsymbol\nabla} H_i= -\chi^{\rm
m}H^2({\boldsymbol\nabla}\chi^{\rm m})/(1+\chi^{\rm m})$,
an apparent contradiction.

A proper analysis of the force should start from the
simple expression of Eq~(\ref{f}). If not, the
contribution of the zero-field pressure, $P(T,\rho_1)$,
in the presence of a temperature and concentration
gradient, needs to be included. In addition, one must
include higher order terms in the density, take
$\tilde\chi= \alpha\rho+\beta\rho^2$, and $\chi=
\tilde\chi/(1-\tilde\chi)\approx\alpha\rho +
(\beta+\alpha^2)\rho^2$. Inserting these respectively
into Eq~(\ref{87}) and (\ref{87a}), we find, for both
cases and a constant $B$-field,
\begin{equation}
{\bf f}^{\rm P}= {\textstyle\frac{1}{2}} B^2
{\boldsymbol\nabla} [\beta\rho^2].
\end{equation}
Assuming $\tilde\chi\sim\rho$ or $\chi\sim\rho$ to hold
strictly is equivalent to taking $\beta=0$ or $\beta=
-\alpha^2$, resulting respectively in ${\bf f}^{\rm P}=0$
and ${\bf f}^{\rm P}=-\chi B^2\nabla\chi$, or
equivalently, $M_i{\boldsymbol\nabla} B_i$ or
$M_i{\boldsymbol\nabla} H_i $.

Next we go on to consider nonlinear constitutive
relations, and shall convince ourselves that the two
Kelvin force expressions Eqs~(\ref{888},\ref{88a}) remain
valid -- though (in a generalization of the above
conclusion) only to linear order in the magnetization and
polarization. We insert Eq~(\ref{83a}) in Eq~(\ref{nl1}),
and equate $-\partial F_{\rm em}/\partial \rho_\alpha$
with $\int(\partial M_i/\partial \rho_\alpha){\rm d}B_i$,
to yield
\begin{equation}\label{nl2}
{\bf f}^{\rm P}=M_i{\boldsymbol\nabla} B_i+
{\boldsymbol\nabla}\int( \rho_\alpha\partial M_i/\partial
\rho_\alpha-M_i){\rm d}B_i.
\end{equation}
Assuming that $\rho_\alpha\partial M_i/\partial
\rho_\alpha=M_i$ at constant $B$ -- equivalent to
$\rho_\alpha\partial\tilde\chi^{\rm m}/\partial
\rho_\alpha=\tilde\chi^{\rm m}$ for {\sc lcr} --- the
integral vanishes, and we retain Eq~(\ref{88a}). If we
start from $\tilde F_{\rm em}=-\int{\bf B}\cdot{\rm
d}{\bf H}$ to evaluate
\begin{eqnarray} {\bf f}^{\rm P}&=&(\partial
\tilde F_{\rm em}/\partial T){\boldsymbol\nabla} T
-\rho_\alpha{\boldsymbol\nabla}(\partial\tilde F_{\rm
em}/\partial\rho_\alpha)\\ &=&{\boldsymbol\nabla}[\tilde
F_{\rm em}-\rho_\alpha\partial\tilde F_{\rm em}/\partial
\rho_\alpha]+B_i{\boldsymbol\nabla} H_i\\&=&
M_i{\boldsymbol\nabla} H_i+
{\boldsymbol\nabla}{\textstyle\int}( \rho_\alpha\partial
M_i/\partial \rho_\alpha-M_i){\rm d}H_i,
\end{eqnarray}
we find ${\bf f}^{\rm P}=M_i{\boldsymbol\nabla} H_i$ if
$\rho_\alpha\partial M_i/\partial \rho_\alpha=M_i$ for
constant $H$.

\section{Experiments}\label{exp}
Having been derived from thermodynamics and conservation
of total energy, momentum and angular momentum, the
theory presented in the preceding chapters is fairly
general, and valid for all magnetizable and polarizable
liquids, from single-component paramagnetic fluid to
ferrofluids, and for their respective electric
counterparts. In the case of ferrofluids, which are
suspensions of ferromagnetic particles, one is tempted to
think that the properties of the particles are important,
especially their orientation (given by the magnetic
moment), and their internal angular momentum. This is
indeed true if a high-resolution, mesoscopic account of
the system is the prescribed goal. However, on a coarser
scale, with many particles per grain and relevant for
most experiments, the theory derived above is quite
adequate, even uniquely appropriate for being not
unnecessarily detailed and complicated.

As in the last chapter, we shall only display the
magnetic terms, accounting for magnetic effects, as the
analogous electric ones are easily obtained via
Eqs~(\ref{rp1}).

\subsection{Field Induced Variation in Densities}\label{denva}
We consider the change in densities, $\Delta\rho$ and
$\Delta\rho_1$, of a magnetizable liquid in equilibrium,
from a region of low (or no) field to one of high field.
In equilibrium, Eqs~(\ref{Equil}) holds. For a
one-component liquid, they reduce to ${\boldsymbol\nabla}
T=0$ and
\begin{equation}\label{89+}
{\boldsymbol\nabla}\xi=\left[\frac{\partial\xi}{\partial\rho}
\right]_{H}\!\!\!\!{\boldsymbol\nabla}\rho
+\left[\frac{\partial\xi} {\partial
H}\right]_\rho\!\!{\boldsymbol\nabla} H=-g{\bf e_z},
\end{equation}
where the chemical potential has been taken as a function
of $\rho, T$ and the field magnitude $H$. It is useful to
rewrite the two thermodynamic derivatives, the second as
$[\partial\xi/\partial H]_\rho$ $=-[\partial B/\partial
\rho]_{H}= -[\partial M/\partial\rho]_{H}=
-H[\partial\chi^{\rm m}/\partial\rho]$, where the last
equal sign is valid only for {\sc lcr}. The first may be
approximated: Without field, the inverse isothermal
compressibility $\rho^2[\partial\xi/\partial\rho]_T=$ $
\rho[\partial P/\partial\rho]_T\equiv$ $\kappa_T^{-1}$ is
usually a large enough quantity that one may neglect the
field related corrections, which is $-\frac{1}{2}(\rho
H)^2(\partial^2\chi^{\rm m}/\partial\rho^2)$ for {\sc
lcr}. Employing this approximation and assuming {\sc
lcr}, we integrate Eq~(\ref{89+}) to yield the variation
in density in response to the gradient of magnetic field
strength and the gravitational potential,
\begin{equation}\label{89a}
\mbox{\boldmath$\Delta$}\rho= \rho^2\kappa_T
\mbox{\boldmath$\Delta$}
({\textstyle\frac{1}{2}}H^2\partial\chi^{\rm
m}/\partial\rho-gz),
\end{equation}
where the boldfaced $\mbox{\boldmath$\Delta$}$ denotes
(as before) the difference of any quantity at two
different points in the liquid,
$\mbox{\boldmath$\Delta$}A\equiv A_2-A_1$. The electric
terms may be obtained, as usual, according to
Eqs~(\ref{rp1}). (Frequently, this effect --- referred to
as the electro- or magnetostriction --- is calculated
using the ponderomotive force of section \ref{pond}. The
above calculation shows that there is no difficulty at
all to avoid the ponderomotive force, and the ambiguity
associated with its notation.) Electrostriction has been
verified~\cite{Ha}, using the refractive index to measure
the density change.

If the fluid has two components -- such as when it is a
solution or suspension -- we need to consider both
chemical potentials, ${\boldsymbol\nabla}\xi_1=0$ and
${\boldsymbol\nabla}\xi=$ $-g{\bf e_z}$. More
conveniently, however, in the incompressible limit of
section~\ref{incomp}, we may consider Eqs~(\ref{equil})
alone,
\begin{eqnarray}\label{89++}
{\boldsymbol\nabla}\xi^{\rm in}_1=\left[
\frac{\partial\xi^{\rm in}_1}{\partial\rho_1}
\right]_{H}\!\!\!\!{\boldsymbol\nabla}\rho_1
+\left[\frac{\partial\xi^{\rm in}_1} {\partial
H}\right]_{\rho_1}\!\!{\boldsymbol\nabla} H=-g\gamma{\bf
e_z},\quad\\ \!\!{\boldsymbol\nabla}\rho_1 =[
\partial\xi^{\rm in}_1/\partial\rho_1]^{-1}_{H}
\left([\partial M/\partial\rho_1]_{H}
\!\!{\boldsymbol\nabla} H-g\gamma{\bf e_z}\right).\quad
\label{89+++}
\end{eqnarray}
(Note that $\partial M/\partial\rho_1$ is connected to
$\partial\xi^{\rm in}_1/\partial H$ via a Maxwell
relation.) To our knowledge, field-induced equilibrium
variation in the solute or particle density has not  yet
been measured in any two-component liquids. This is
unfortunate especially in ferrofluids, where the
variation in particle density should be rather
pronounced. Quantitatively, this experiment yields the
thermodynamic derivative $\partial\xi^{\rm in}_1/
\partial\rho_1 =[(\rho_1)^2 \kappa_{\rm os}]^{-1}$, with
$\kappa_{\rm os}$ the osmotic compressibility. Since this
is a diagonal derivative, its significance in
characterizing the ferrofluid ranks with that of the
compressibility, specific heat and magnetic
susceptibility.

Let us estimate the magnitude of this effect: Writing
Eq~(\ref{89+++}) as ${\boldsymbol\Delta}\rho_1/\rho_1=
\kappa_{\rm os}(\rho_1\partial\chi^{\rm
m}/\partial\rho_1) {\boldsymbol\Delta}
(\mu_0{\textstyle\frac{1}{2}}\hat H^2)$, we approximate
$(\rho_1\partial\chi^{\rm m}/\partial\rho_1)\approx1$,
estimate $\kappa_{\rm os}\approx10^{-3}/$Pa, and find
${\boldsymbol\Delta}\rho_1/\rho_1\approx0.1$ for $\hat
B=10^{-2}$T. [The value for $\kappa_{\rm os}$ is obtained
by considering a ferrofluid with 10\% of its volume
occupied by magnetic particles of the radius $r=10$nm, so
the particle density is $n_1=0.1/(4\pi r^3/3)$.
Approximating these particles as ideal gas, the inverse
osmotic compressibility $\kappa^{-1}_{\rm os}$ is equal
to the osmotic pressure, $P_{\rm os}=n_1k_{\rm B}T$, so
$\kappa_{\rm os}=10^{-3}/$Pa if $T=300$K.] Compare this
with tiny change of the total density,
$\Delta\rho/\rho=5\cdot10^{-8}$ at the same field, cf
Eq~(\ref{89a}), a result of the small total
compressibility, $\kappa=5\cdot10^{-10}/$Pa.

\subsection{Current Carrying Vertical Wire}\label{vccw}

We consider a vertical wire that goes through a dish
filled with ferrofluid. Feeding the wire with electric
current will drag the ferrofluid toward the wire (located
at $r=0$ and along $z$ in cylindrical coordinates). The
surface of the ferrofluid column is given by $z(r)$, with
$z$ at which the radius diverges (obtained by
extrapolation) as the origin, ie $z(\infty)=0$. The
boundary condition Eq~(\ref{K}) is evaluated for two
points, $z(r)$ and 0. Because the magnetization vanishes
and the curvature radii diverge at 0, the attendant
result is $K=P_{\rm atm}$. Inserting this into the
boundary condition at $z(r)$, we obtain
\begin{eqnarray}\label{smb}
{\textstyle\int_{\rm Eq}}M_i{\rm d}H_i
=\alpha(R_1^{-1}+R_2^{-1})+\bar\rho gz,
\end{eqnarray}
because $M_n=0$ for the given geometry. The integral is
to be evaluated for $H=J/(2\pi r)$, and at constant $T,
\xi_1$ if equilibrium has time to be established. For
times much briefer after the current has been applied, we
need to substitute $\rho$ for $\bar\rho$, and $\int_{\rm
uni}$ for $\int_{\rm Eq}$, and evaluated the integral at
constant $T$, $\rho_1$, cf the discussion leading to
Eq~(\ref{sn2}). Neglecting the surface tension,
$\alpha=0$, and taking the subscript $\int_{\rm uni}$
lead to the result given in~\cite{rz}.

In the spirit of the last paragraph of
section~\ref{surfF}, we remark that Eq~(\ref{smb}) may be
considered as an expression for force equilibrium,
between gravitation, surface tension and the magnetic
surface force.

For {\sc lcr},  $M_i=\chi^{\rm m}H_i$, the left hand side
reduces to $\frac{1}{2}\chi^{\rm m}H^2$, and we are left
with a quadratic, hyperbolic profile of the interface,
$8\pi^2\bar\rho$ $gz=$ $J^2\chi^{\rm m}/r^{2}$ if the
surface tension $\alpha$ is neglected. The effect of
$\alpha$ is more important for weak currents, $J$ small.
It may be neglected in any case for $z\to0$, where both
curvature radii are large enough to be ignored. For $z$
large, one curvature radius is simply $r$, and the other
$\infty$. So this part of the ferrofluid column is
accounted for by $8\pi^2$ $(\bar\rho gz+$ $\alpha/r)=$
$J^2\chi^{\rm m}/r^{2}$, with the term $\sim r^{-2}$
being asymptotically ($r\to0$) the dominant one. In
between, where the actual bend from the horizontal to the
vertical takes place, both curvature radii are finite and
need to be included for an understanding of the surface
-- note, however, that they have different signs.

\subsection{Hydrostatics in the Presence
of Fields}\label{levi}

In a system of two connected tubes, with only the second
subject to a magnetic field, we expect the ferrofluid
column to be higher in this tube, as ferrofluid is
attracted to the region of stronger fields, see Fig 3. To
calculate the level difference, we employ the boundary
condition Eq~(\ref{K}) for the (flat) liquid-air
interface in both tubes. Since the field vanishes in the
first tube (denoted as 1), the boundary condition simply
states $P_{\rm atm}=K-\bar\rho gz_1$. Inserting this into
the boundary condition for tube 2, we obtain
\begin{equation}\label{302}
\bar\rho g(z_2-z_1)={\textstyle\int_{\rm Eq}}M_i{\rm
d}H_i+{\textstyle\frac{1}{2}}M_n^2.
\end{equation}
Note again the force balance between the electromagnetic
surface force Eq~(\ref{emsf2}), operational at the
surface in tube 2, and the gravitational force from the
disparity in height. (As in section \ref{MSE}, the
integral is to be evaluated at constant $T$ and $\xi_1$
if equilibrium has time to be established. For much
briefer times after the field has been applied,
$\bar\rho$ is to be replaced by $\rho$, and the integral
by $\int_{\rm uni}$, evaluated at constant $T,\rho_1$.)

\begin{figure}
\begin{center}
\includegraphics[width=0.5\columnwidth,angle=0]
{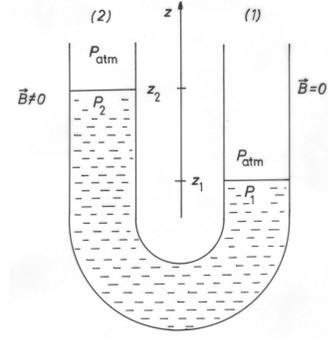}
\end{center}
\caption{A U-tube filled with ferrofluid with only the
left arm subject to a field. The pressure $P_{\rm atm}$
outside the ferrofluid is constant. Depending on the
orientation of the field, the  level difference $z_2-z_1$
is given by Eqs~(\ref{123y}) and
(\ref{124y}).}\label{Fig3}
\end{figure}

If the field is either predominantly tangential to the
liquid surface ($H=H_t$, $B=B_t$ and $H_n, B_n=0$), or
predominantly normal to the liquid surface ($H=H_n$,
$B=B_n$ and $H_t$, $B_t=0$), we have respectively,
\begin{eqnarray}\label{123y}
g\bar\rho(z_2-z_1)={\textstyle\int_{\rm Eq}}M_i{\rm
d}H_i=\textstyle{\frac{1}{2}}H^2\chi^{\rm m}\\
\label{124y} g\bar\rho(z_2-z_1)={\textstyle\int_{\rm
Eq}}M_i{\rm d}B_i= \textstyle{\frac{1}{2}}H^2 \chi^{\rm
m}\mu,
\end{eqnarray}
with the last equal signs in both equations only valid
for {\sc lcr}.

Next we consider the quantity that a pressure gauge
measures in a ferrofluid exposed to a magnetic field. As
emphasized, it is not the ordinary pressure, yet as it
will give some value, the question is what this is. Think
of the gauge as an enclosed volume of air, at the
pressure $P_{\rm atm}$, see Fig 4. One side of this
volume is an elastic membrane, which is displaced if the
external stress tensor deviates from the internal one. A
finite displacement $d$ stores up the elastic energy
$kd^2/2$ per unit area of the membrane. (Take the
membrane to be stiff, ie $k$ large and $d$ small, then we
need not worry about the pressure change inside.) The
elastic energy implies a force density $kd$, rendering
the boundary condition across the membrane as
$\triangle\Pi^{\rm tot}_{nn}=kd$, or via Eq~(\ref{K}),
$K- P_{\rm atm}+ {\textstyle\int_{\rm Eq}}M_i{\rm d}H_i+
\textstyle\frac{1}{2}M_n^2- \bar\rho gz= kd$. ($d$ is
taken to be positive when the membrane protrudes into the
gauge.)
\begin{figure}
\includegraphics[width=0.8\columnwidth,angle=0]
{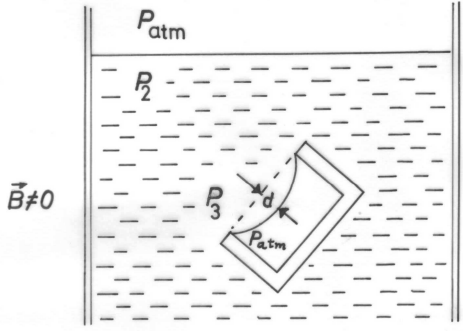} \caption{A vessel filled with ferrofluid is
subject to a field. The pressure inside the ferrofluid
changes with height and the field strength, in response
to which the elastic membrane of a pressure gauge (right
bottom) yields. The corresponding displacement $d$ is
given by Eq~(\ref{303}).}\label{Fig4}
\end{figure}
We have $d=0$ as long as the external stress tensor is
the same as the internal one, ie in the atmosphere down
to the liquid surface of tube 2, and also just below the
surface -- take this as point 2. Moving down the liquid
column, to an arbitrary point 3, the membrane moves the
distance $d$ to maintain force equilibrium. Employing the
above boundary condition for both points, and subtracting
one from the other, we find
\begin{equation}\label{303}
kd=\mbox{\boldmath$\Delta$}({\textstyle\int_{\rm
Eq}}M_i{\rm d}H_i+{\textstyle\frac{1}{2}}M_n^2-\bar\rho
gz),
\end{equation}
with $\mbox{\boldmath$\Delta$}A \equiv A_3-A_2$. Note
$\mbox{\boldmath$\Delta$}M_n \equiv M^n_3-M^n_2$, where
$M^n_3$ is the magnetization at point 3 normal to the
pressure gauge membrane, and $M^n_2$ the field at point 2
normal to the liquid surface -- both components are not
necessarily parallel. Eq~(\ref{303}) again states a force
balance, between the elastic, magnetic surface and
gravitational force.

The displacement $d$ is the readout of the pressure
gauge: If the field is uniform, and if the membrane of
the gauge is parallel to the liquid surface,
Eq~(\ref{303}) reduce to $\rho g( z_2-z_3)=kd$, the
zero-field hydrostatic relation; otherwise, field
contributions abound -- even if the pressure gauge is
simply rotated at point 3 in the presence of a uniform
field. This amply demonstrates the system's anisotropy. A
further complication is that all fields are the actual
ones, distorted by the presence of the pressure gauge --
though this is an effect that may be minimized.

\subsection{Magnetic O-Rings and Scrap Separation}
\label{scrap}

In this section, we address the physics of some technical
applications: magnetic O-rings, self-levitation and scrap
separation.  Consider scrap separation first. An
inhomogeneous magnetic field which becomes weaker with
increasing height may lift non-polarizable bodies
submerged in ferrofluids off the ground, and hold them at
specific heights which depend on the shape and density of
the bodies.

The calculation is already given in section~\ref{surfF}.
Balancing the gravitation with magnetic force, we have
${\pmb{\cal F}}^{\rm elm}+{\pmb{\cal F}}^{\rm grav}=0$,
where ${\pmb{\cal F}}^{\rm grav}=$ $-V(\rho^{\rm
s}-\rho^{\rm f})g\hat{\bf e}_z$, and ${\pmb{\cal F}}^{\rm
elm}$ is given by Eqs~(\ref{emsf2}, \ref{emsf}), and
especially by Eqs~(\ref{emsf3}, \ref{emsf4}) if the solid
is completely nonmagnetic. (Again, depending on the time
scale, the integral ${\textstyle\int_{\rm uni}}$ may be
the appropriate one, cf section~\ref{MSE}.)

Because scrap separation is an equilibrium
pheno\-me\-non, we may compare energies instead of
computing forces. This is a simpler and more qualitative
approach to understand the behavior of polarizable
systems. Consider for instance the fact that ferrofluid
is attracted to the region of high magnetic field. Take
the field $B$ as given, the field energy per unit volume
is $B^2/2$ in vacu\-um (or air), and $B^2/(2\mu)$ in
ferrofluids. Since $\mu>1$ in any paramagnets, the second
expression is smaller. Given the choice, a volume element
of ferrofluid will therefore always occupy a region with
the highest possible field, to reduce the field energy.
Conversely, a small, non-magnetizable object, submerged
in ferrofluid, will on the other hand tend to occupy the
region of lowest field strength. If a difference in
height is involved, all these of course happen only as
long as the gain in field energy is larger than the loss
in gravitational energy levitating the object. (If the
field $H$ is given instead of $B$, we need to consider
$\tilde F_{\rm em}\equiv F_{\rm em}-HB$. And again, it is
larger in vacuum then in the ferrofluid: $-H^2/2>-\mu
H^2/2$.)

\begin{figure}[ht]
\includegraphics[width=0.7\columnwidth,angle=-1]
{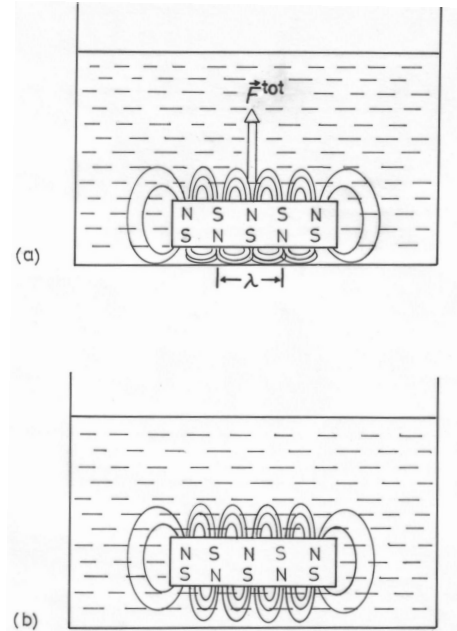} \caption{Self-levitation of a magnetized body
submerged in ferrofluid.}\label{Fig5}
\end{figure}
Similarly, a permanently magnetized body submerged in
ferrofluid tends to collect as much liquid around itself
as possible, in the space occupied by its stray field --
even at the price of levitating itself off the bottom, a
phenomenon that is usually referred to as
``self-levitation"~\cite{rz}. Frequently, the magnetized
body consists of a periodic array of north and south
poles, with periodicity $\lambda$, so the stray field
extends one or two $\lambda$ into the ferrofluid.
Levitated approximately that far from the bottom, the
body will usually have reached its equilibrium position,
as no further gain in field energy may be achieved by
pushing more ferrofluid below it, and levitating itself
yet higher, Fig 5.

Generally speaking, although energetic considerations are
a useful heuristic tool, the actual minimization of total
energy or free energy is often rather cumbersome, as the
field does change considerably when a volume element of
ferrofluid is displaced. For quantitative calculations,
it is therefore frequently more convenient to consider
boundary conditions involving the Maxwell stress tensor,
as we have done throughout this review article.

This also pertains to magnetic O-rings, which may be
found in most computer hard disk drives and are perhaps
the most widely deployed ferrofluid device. In these,
some ferrofluid is positioned as an O-ring between a
highly permeable rotating shaft and the pole of a
permanent magnet, Fig 6. Serving as a pressure seal, it
enables the rotary component to work in vacuum. In
contrast to the above circumstances, we are dealing with
a metastable state here, as it is always energetically
more favorable to remove the ferrofluid seal first, have
the pressure equalized, and then return the seal to its
original place at the poles. So the quantity of interest
is the lowest energy barrier that must be overcome, which
in any realistic problems is a quantity notoriously
difficult to find. The correct force balance, on the
other hand, is easy to write down:
\begin{equation}\label{ml4}
A\mbox{\boldmath$\Delta$}P=\mbox{\boldmath$\Delta$}
\int({\textstyle \frac{1}{2}}M^2_n+ {\textstyle\int_{\rm
Eq}}M_k{\rm d}H_k){\rm d}{\bf A},
\end{equation}
expressing the balance between the difference in the
electromagnetic surface force [cf Eq~(\ref{emsf2})] on
the two free surfaces (area A) of the ferrofluid and the
difference in external pressure. The equation is obtained
by considering the boundary condition
$\triangle\Pi_{nn}^{\rm tot}=0$, for both free surfaces,
call them 1 and 2. Use Eq~(\ref{301}), forget
gravitation, and remember that $P$ is different for 1 and
2 to arrive at Eq~(\ref{ml4}), with
$\mbox{\boldmath$\Delta$} P=P_0(1)-P_0(2)$.
\begin{figure}
\includegraphics[width=0.99\columnwidth,angle=0]
{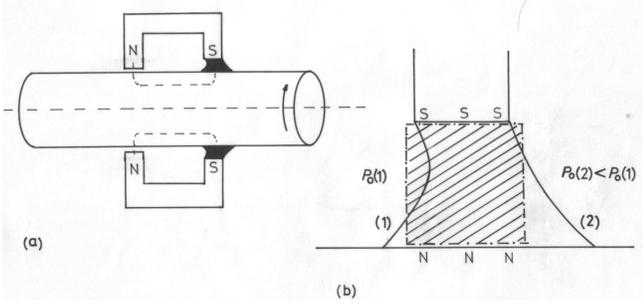} \caption{Magnetic pressure seal: (a) principle
and (b) enlarged displacement of the ferrofluid
plug.}\label{Fig6}
\end{figure}

The field is strongest in the middle of the O-ring and
decays towards both ends. If $\mbox{\boldmath$\Delta$}P$
were zero, the force ${\bf F}^{\rm elm}$ would be the
same on both surfaces, and the ferrofluid stays in the
middle of the O-ring. Increasing the pressure on the left
(surface 1) pushes the ferrofluid towards right, such
that surface 1 is in the region of higher, and surface 2
is in the region of weaker, fields. Equilibrium is
achieved when the difference in ${\bf F}^{\rm elm}$
balances $\mbox{\boldmath$\Delta$} P$. The strongest
pressure difference maintainable is when one surface is
at the region of highest field, and the other is
field-free. Assuming for simplicity that the magnetic
field is predominantly tangential, and that {\sc lcr}
holds, we have $\mbox{\boldmath$\Delta$}
P=\frac{1}{2}\chi^{\rm m}H^2$. With $\chi^{\rm
m}\approx1$, $H^2= \mu_0\hat H^2$, and taking $\hat H$ as
of order $10^5$A/m, this pressure difference is about
$10^5$N/m$^2$, approximately the atmospheric pressure.

\subsection{Elliptical Deformation of
Droplets}\label{elli} A droplet of ferrofluid exposed to
an external magnetic field along $\hat{\bf e}_y$ is
distorted, cf eg~\cite{bacri}. If the field is not too
strong, it turns from a sphere of radius $r$ to an
ellipsoid of essentially the same volume, $y^2/a^2+
(x^2+z^2)/b^2=1$, with $a, b$ the semimajor and semiminor
axes, $b<a$, see Fig 7. Given a uniform applied field,
$B_{\rm ex}$, the internal field of a magnetizable
ellipsoid is also uniform, and along the applied field,
even for nonlinear constitutive relation, cf~\S~8 of
\cite{LL8}. Once these facts are assumed, it is fairly
simple to calculate the distortion, even nonlinearly, as
a function of the field, because again, only the boundary
condition, $\triangle\Pi_{nn}^{\rm tot}=
\alpha(R_1^{-1}+R_2^{-1})$ from Eqs~(\ref{54}), needs to
be evaluated, at the points $(x,y,z)=(0,a,0)$ and
$(b,0,0)$. They are respectively
\begin{equation}\label{206}
\triangle\Pi^{\rm tot}_{yy}=\alpha(2a/b^2),\quad
\triangle\Pi^{\rm tot}_{xx}= \alpha(1/b+ b/a^2).
\end{equation}
Employing Eq~(\ref{K}) to evaluate the respective left
side of these two equations, we find $\triangle\Pi^{\rm
tot}_{yy}- \triangle\Pi^{\rm tot}_{xx}= \frac{1}{2}M^2$,
because all fields are constant within the ellipsoid, and
because $M_n=M$ at $(0,a,0)$, $M_n=0$ at $(b,0,0)$. (The
extent of the droplet is presumed small enough for us to
neglect the gravitation.) Taking the difference also of
the right hand side, we arrive at
\begin{eqnarray}\label{207}
\textstyle\frac{1}{2}M^2= (\alpha/r)
(2\eta^{-2/3}-\eta^{-1/6} -\eta^{5/6}),
\end{eqnarray}
where the parameter $\eta$ is related to the eccentricity
$e$: $\eta=1-e^2= (b/a)^2$. (With $r^3=ab^2$, we have
$a=r\eta^{-1/3}$, $b=r\eta^{1/6}$.) This result is the
same as that given in~\cite{Berg}.  Both sides of this
equation may be appro\-ximated independently: For small
eccentricity, the right side reduces to $(\alpha/r)2e^2$,
while the  left side reduces to $\textstyle\frac{1}{2}
(\mu -1)^2H^2$ for {\sc lcr}. In addition, we have
$H_{\rm ex}/H= 1+n(\mu-1))$, where $H_{\rm ex}$ is the
applied field, and $n=(1-2e^2/5)/3$ for small
eccentricity, cf \S 8 in \cite{LL8}.
\begin{figure}
\includegraphics[width=0.6\columnwidth,angle=0]
{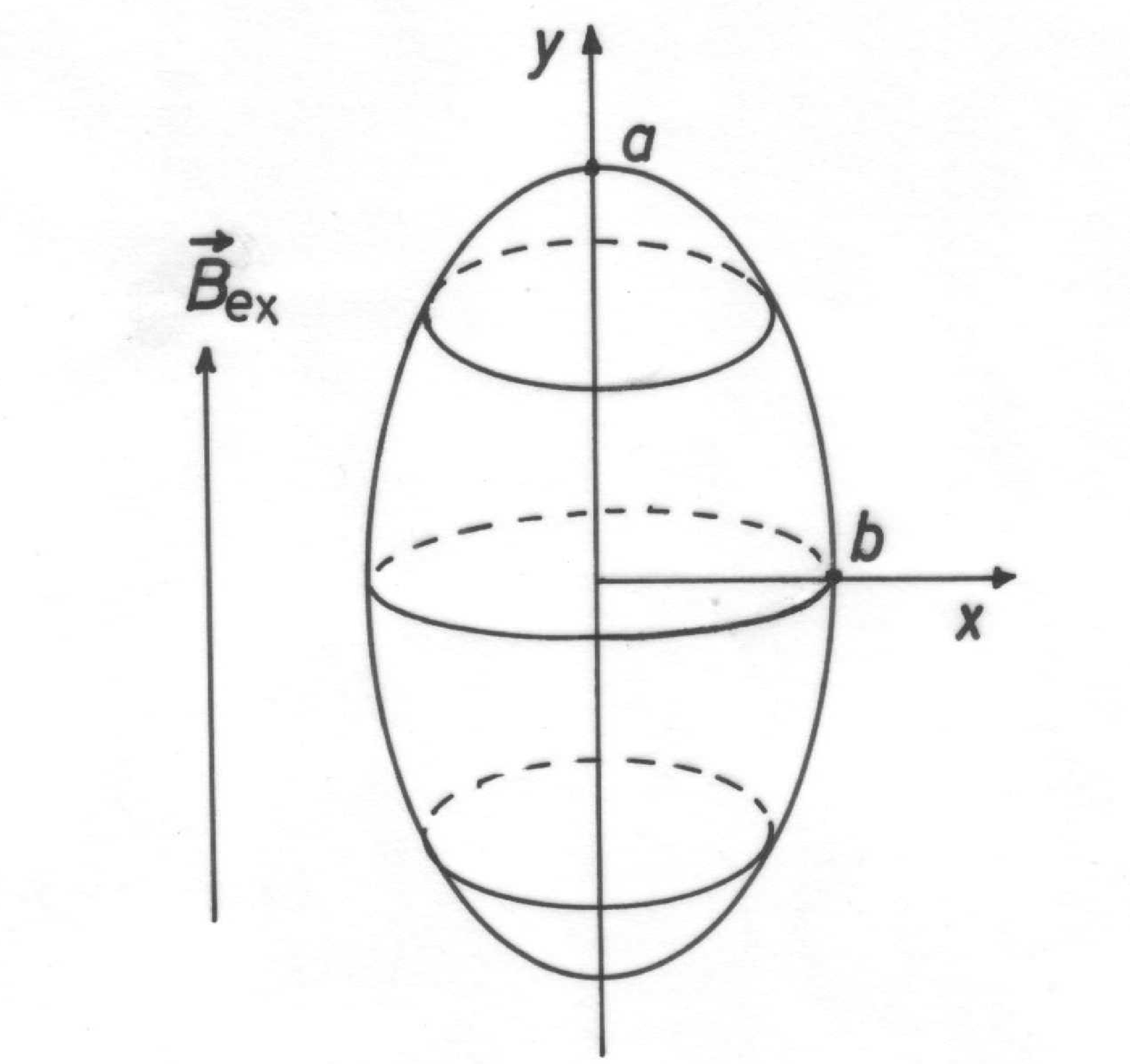} \caption{A freely suspended ferrofluid droplet
elongates along the field, the extent of which is given
by Eq~(\ref{207}).}\label{Fig7}
\end{figure}

\section{Appendix}\label{app}
The validity of Eq~(\ref{40}) is shown here directly by
transforming the rest frame expression. More
specifically, we demonstrate $\partial
(u-v\mbox{\boldmath$g$}^{\rm tot})/\partial {\bf D}={\bf
E}_0$, holding $s$, $\rho_\alpha$, $\mbox{\boldmath$v$}$,
and $\bf B$ constant. We start with

\begin{equation}
u=u_0({\bf D_0\to D,\ B_0\to
B})+\textstyle\frac{1}{2}\rho v^2.
\end{equation}
This pleasingly simple expression is a result of the
accidental cancellation of the terms from the
Galilean-Lorentz transformation with that of the Tailor
expansion,
\begin{eqnarray}
u({\bf D}, {\bf B})=u_0({\bf D}_0,{\bf
B}_0)+2\mbox{\boldmath$v$}
\cdot\mbox{\boldmath$g$}_0^{\rm
tot}+\textstyle\frac{1}{2}\rho v^2\nonumber\\ =u_0({\bf
D}_0,{\bf B}_0)+2\mbox{\boldmath$v$} \cdot({\bf E\times
H})/c +\textstyle\frac{1}{2}\rho v^2\nonumber\\ =u_0({\bf
D},{\bf B})+\textstyle\frac{1}{2}\varrho v^2. \nonumber
\end{eqnarray}
Assuming {\sc lcr}, or
$u=\textstyle\frac{1}{2}(D^2/\epsilon+\rho v^2)$ with
${\bf D}_0=\epsilon{\bf E}_0$, we have
\begin{eqnarray}\label{yi}\nonumber
&&u-\mbox{\boldmath$v$}\cdot\mbox{\boldmath$g$}^{tot}
=\textstyle\frac{1}{2}(D^2/\epsilon-\rho
v^2)-\mbox{\boldmath$v$} \cdot({\bf E\times H})/c\\
&&=\textstyle\frac{1}{2}(D^2/\epsilon-\rho
v^2)-\mbox{\boldmath$v$} \cdot({\bf D}\times{\bf
H})/c\epsilon+{\cal O}(v^2/c^2),\quad
\end{eqnarray}
and deduce
\begin{eqnarray}
&&\partial(u-\mbox{\boldmath$v$}\cdot
\mbox{\boldmath$g$}^{tot}) /\partial {\bf D}\nonumber\\
&&=({\bf D}+\mbox{\boldmath$v$}\times{\bf
H}/c)/\epsilon={\bf D}_0/\epsilon={\bf E}_0.
\end{eqnarray}
Higher order terms (such as one $\sim D^4$ in the energy
$u$) do not invalidate this result. The terms in the
magnetic field behave analogously.

In Eq~(\ref{yi}), the explicit form of
$\mbox{\boldmath$g$}^{\rm tot}$ in the lab frame was
employed to deduce the lab frame energy, Eq~(\ref{40}),
from which then the lab frame energy flux, Eq~(\ref{39}),
is deduced. This may appear as an inconsistency, but is
not, because with the rest frame expression for
$\mbox{\boldmath$g$}^{\rm tot}_0$ given, we already know
that the term $\sim\mbox{\boldmath$v$}$ is from the rest
mass. No detailed information about the energy flux is
necessary here.

\vspace{1.6cm} {\bf Acknowledgment:} We thank Andreas
Engel, Kurt Sturm and Hubert Temmen for comments, and we
are especially grateful to Hanns-Walter M\"{u}ller for
detailed criticisms and suggestions.


\begin{thebibliography}{99}
\bibitem{fd} We shall use the terms ``force/force
density," ``energy/energy density" interchangably. 
\end{thebibliography}
\end{document}